\pdfoutput=1 

\documentclass[prd,aps,preprint,amsmath,nofootinbib,amssymb,eqsecnum,showkeys,tightenlines]{revtex4-1}

\usepackage{amsmath, latexsym, amssymb, hyperref, graphicx, color, multirow}
\usepackage[table]{xcolor}
\usepackage{tabularx}
\usepackage[T1]{fontenc}
\usepackage[capitalize]{cleveref}
\definecolor{nicered}{rgb}{.7,.1,.1}
\definecolor{nicegreen}{rgb}{.1,.5,.1}
\definecolor{darkblue}{rgb}{0,0,.5}
\hypersetup{colorlinks, citecolor=nicegreen,linkcolor=nicered, urlcolor=darkblue}
\usepackage{multirow}
\usepackage{slashed}
\numberwithin{equation}{section}

\newcommand\myeq{\mathrel{\overset{\makebox[0pt]{\mbox{\normalfont\tiny\sffamily rest frame}}}{\rightarrow}}}

\begin{document}
\preprint{KIAS P17141, MI-TH-1774}
\preprint{}

\title{Prospects for discovery and spin discrimination of dark matter in Higgs portal DM models and their extensions 
at 100 TeV $pp$ collider }

\author{Bhaskar Dutta$^a$}
\email{dutta@physics.tamu.edu}
\author{Teruki Kamon$^a$}
\email{kamon@physics.tamu.edu}
\author{P. Ko$^b$}
\email{pko@kias.re.kr}
\author{Jinmian Li$^{b,c}$}
\email{jmli@kias.re.kr}
\affiliation{$^a$ Mitchell Institute for Fundamental Physics and Astronomy, Department of Physics and Astronomy, Texas A\&M University, College Station, TX 77843-4242, USA}

\affiliation{$^b$ School of Physics, Korea Institute for Advanced Study, Seoul 130-722, Korea}

\affiliation{$^c$ Center for Theoretical Physics, College of Physical Science and Technology, Sichuan University, Chengdu, 610064, China}

\begin{abstract}
\noindent
We study the discovery and discriminating prospects of the Higgs portal dark matter (DM) models 
for scalar, fermion and vector DM  and their extensions in proton-proton ($pp$) collisions. 
The $t\bar{t}+$DM associated production in dileptonic final states is considered, 
in which the stransverse mass of two leptons is found to be effective in suppressing the Standard 
Model backgrounds along with the missing transverse energy and the angle between two leptons. The distributions of missing transverse energy and polar angle between two leptons are used for a discrimination of the spin nature of DM.  
For the proposed benchmark points, the discovery/exclusion can be made with an integrated luminosity less than 1 ab$^{-1}$ given a 1\% systematic uncertainty, while the spin discrimination require integrated luminosity of a few O(10) ab$^{-1}$ given a 0.5\% systematic uncertainty. 
The DM phenomenology is also discussed. A consistent DM candidate can be obtained either 
by extending our model where the Higgs portal couples to excited dark states 
that decay into DM, or modifying the coupling form into pseudoscalar. 

\end{abstract}

\maketitle

\section{Introduction}\label{sec:intro}

The existence of dark matter (DM) has been confirmed by astrophysical observations, 
such as galaxy rotation curve~\cite{Corbelli:1999af}, bullet cluster collision~\cite{Clowe:2003tk}, 
cosmic microwave background (CMB) anisotropy~\cite{0067-0049-208-2-20}. 
A precise measurement by the Plank satalitte~\cite{Ade:2015xua} indicates that 26\% of the 
total energy of our universe is made of nonbaryonic  DM.   
Even with null results from underground direct detection experiments and 
lepton/hadron colliders, there are its elusive hints 
at a few indirect detection experiments in the space recently, e.g. Fermi-LAT galactic center excess~\cite{Daylan:2014rsa},  AMS02 anti-proton excess~\cite{Aguilar:2016kjl} and DAMPE 
electron/positron anomaly~\cite{Ambrosi:2017wek}. However, interpretations in terms of
DM annihilation/decay are rather ambiguous because of astrophysical uncertainties. 

In contrast to those indirect detection experiments, probing the DM signals at colliders could elucidate the particle physics properties of DM (e.g., couplings, spins) without suffering from astrophysical uncertainties.
In the framework of a simplified model, where the DM is neutral under the Standard Model (SM) gauge group and interacting with the SM particles via the portal of a single mediator, many studies~\cite{Haisch:2013fla,Haisch:2016gry,Haisch:2013ata,Buckley:2014fba,Harris:2014hga,
Haisch:2015ioa,Backovic:2015soa,Buckley:2015ctj} are devoted to identify the spin (CP property) of the 
mediator and resolve the coupling between the mediator and SM particles. 
However, all those searches mainly focus on the properties of the mediator, and 
the DM information is usually unavailable.  Because the DM is dominantly produced by the decay 
of the on-shell mediator in the simplified model, those visible final states do not carry any 
useful information on the particle physics nature of the DM. 
Many other studies~\cite{Cotta:2012nj,Crivellin:2015wva,Belyaev:2016pxe} consider the characterization of DM spin and its coupling to SM particles in the framework of DM effective field 
theory (EFT). The DM EFT, which is mainly advantaged by its generality, may not be an appropriate 
description of an UV-completion at the colliders~\cite{Buchmueller:2013dya,Busoni:2013lha,Busoni:2014sya,Busoni:2014haa,Baek:2015lna,Arcadi:2017kky}.  
The DM characterizations in simplified models of some UV-completions are studied until recently. Refs.~\cite{Ko:2016xwd,Kamon:2017yfx} studied the DM spin discrimination in the Higgs portal DM models at future electron-positron collider.  It was also found that the DM spin can also be revealed at LHC through its radiative corrections to the Drell-Yan process~\cite{Capdevilla:2017doz} and spectral decomposition of the mono-jet signature~\cite{Bae:2017ebe}. 
Reference~\cite{Baum:2017kfa} shows that the DM properties can be determined by combining both the direct detection and collider signals. 

In this paper, as proceeding to our works in Refs.~\cite{Ko:2016xwd,Kamon:2017yfx}, we study 
the discovery and spin discriminating prospects of Higgs portal DM models with scalar DM, fermion 
DM and vector DM at future hadron colliders. It was found in our previous study~\cite{Ko:2016ybp} 
that the Higgs portal DM model is well below the current sensitivity of LHC, due to its small scalar 
mixing angle as required by the SM Higgs precision measurement. 
Even the high luminosity LHC would only be able to probe some portion of the parameter space 
in the Higgs portal DM models. 
On the other hand, 100 TeV colliders~\cite{Mangano:2016jyj,Tang:2015qga} have been proposed 
to explore directly a much larger region of the landscape of new physics models, such as FCC-hh 
and SppC. The target integrated luminosity can reach as high as $25$ ab$^{-1}$~\cite{Zimmermann:2016puu}. So we conduct our studies at 100 TeV proton-proton collider in this work. 
At the LHC, the DMs in the Higgs portal models are usually searched through the mono-jet signature, due to its largest production cross section. However, recent experimental results~\cite{Sirunyan:2017hci,Sirunyan:2017leh} show that the $t\bar{t} +$ DM associated production has a comparable sensitivity with the mono-jet channel if the SM fermions-mediator couplings are proportional to 
Yukawa couplings. The $t\bar{t} +$ DM production will be benefited much more than the mono-jet channel 
by increasing the collision energy from 14 TeV to 100 TeV collider. Much smaller energy fraction is 
required from the parton distribution function of proton,  which results in a dramatically
increased production cross section. Furthermore,  the $t\bar{t}+$ DM signature provides useful 
observables for the DM spin discrimination. 

This paper is organized as follows. In Sec.~\ref{sec:model}, the models are introduced and some possible DM spin discrimination variables are proposed. Sec.~\ref{sec:collider} details the collider searches for the DM and the strategy for the spin discrimination based on a few benchmark 
points. For completeness, the DM phenomenology are studied in Sec.~\ref{sec:dm}, where two 
possible solutions to evade the stringent results by DM direct detection experiments are discussed. 
We summarize the work in Sec.~\ref{sec:concl}. 

\section{Models and signals} 
\label{sec:model} 

In this work, we will consider minimal Higgs portal DM models for scalar, fermion and vector DM particles, which are required to conserve the SM gauge symmetry and renormalizability. 
Since the models have been discussed in Refs.~\cite{Baek:2011aa,Baek:2012se,Kamon:2017yfx},
we simply list the interaction Lagrangians for three types of Higgs portal DM models relevant to the collider phenomenology.   
\begin{align}
\mathcal{L}^{\text{int}}_{\text{SDM}} &=  - h~ \left( \sum_f \frac{m_f}{v_h} \bar{f} f - \frac{2 m^2_W}{v_h} W^+_\mu W^{- \mu} - \frac{m^2_Z}{v_h} Z_\mu Z^\mu \right)   - \lambda_{HS} v_h~ h S^2,  \label{eq:lsdm} \\
\mathcal{L}^{\text{int}}_{\text{FDM}} &= - \left(H_1 \cos \alpha + H_2 \sin \alpha \right) \left( \sum_f \frac{m_f}{v_h} \bar{f} f 
- \frac{2m^2_W}{v_h} W^+_\mu W^{-\mu} - \frac{m_Z^2}{v_h} Z_\mu Z^\mu  \right) \nonumber \\ 
     & + g_\chi \left(H_1 \sin \alpha - H_2 \cos \alpha \right) ~ \bar{\chi} \chi ~,~ \label{eq:lfdm} \\ 
\mathcal{L}^{\text{int}}_{\text{VDM}} &= - \left(H_1 \cos \alpha + H_2 \sin \alpha \right) \left( \sum_f \frac{m_f}{v_h} \bar{f} f 
- \frac{2m^2_W}{v_h} W^+_\mu W^{-\mu} - \frac{m_Z^2}{v_h} Z_\mu Z^\mu  \right) \nonumber \\ 
     & - \frac{1}{2} g_V m_V \left(H_1 \sin \alpha - H_2 \cos \alpha \right) ~V_\mu V^\mu ~,~ \label{eq:lvdm}
\end{align} 
where the subscripts SDM, FDM and VDM denote scalar DM, fermion DM and vector DM, respectively. 
The important fact is that there is only one scalar mediator, i.e. the SM Higgs ($h$), in the SDM case, 
while there is an extra singlet scalar in FDM and VDM cases because of the full SM gauge 
symmetry and $U(1)$ dark gauge symmetry. This singlet scalar shall mix with the SM Higgs,  
giving two scalar mediators in the mass eigenstates ($H_1$, $H_2$) with a mixing angle ($\alpha$). 

\begin{figure}[htb]
\begin{center}
\includegraphics[width=0.35\textwidth]{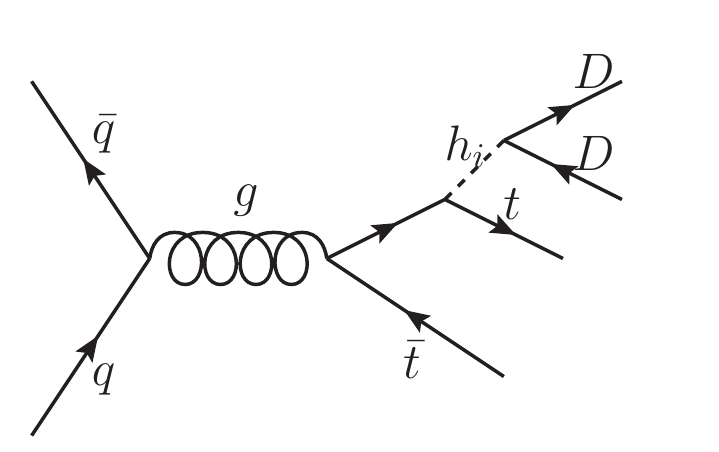}
\includegraphics[width=0.3\textwidth]{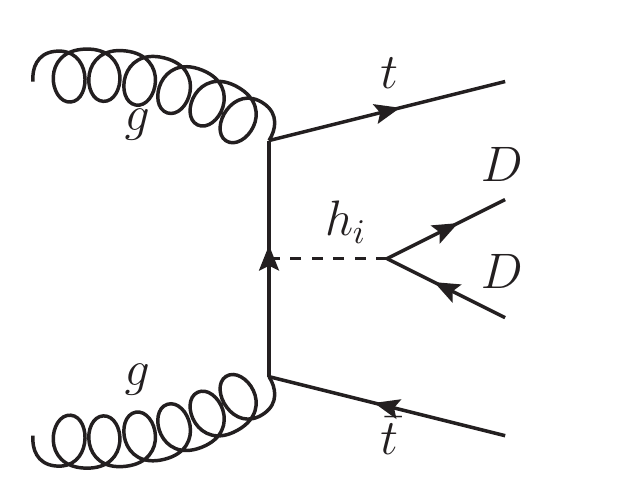}
\includegraphics[width=0.3\textwidth]{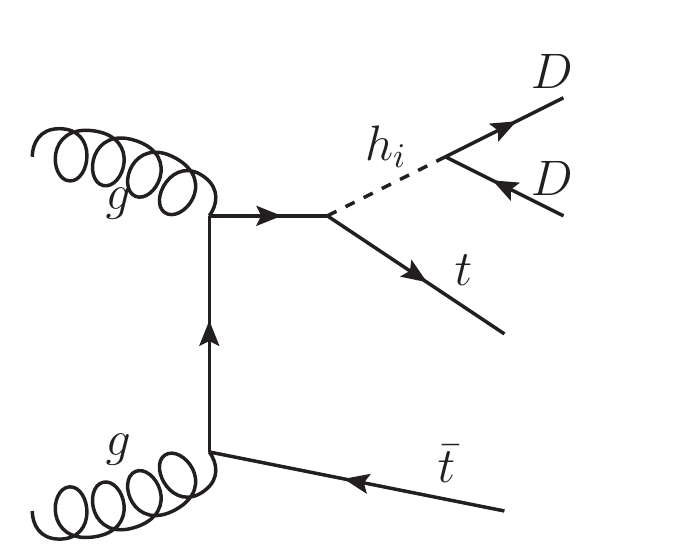}
\end{center}
\caption{\label{fig:sigs} Dominant $t\bar{t}+$DM associated production processes. The DM particle is denoted as $D$ for generic case without specifying the spin.}
\end{figure}

For the DM production with top quark pair, the dominant ones are presented in the Fig.~\ref{fig:sigs}. 
To understand the main kinematic features for each DM spin, it will be illustrative to present the differential production cross sections with respect to the variable $t \equiv m^2_{DD} = (p_{D_1} +p_{D_2})^2$. 
Because of the scalar nature of the mediators, the differential cross section can be factorized into the off-shell mediator production with mass $t$ and its decay: 
\begin{align}
\frac{d \sigma_{\text{SDM}}}{d t} & \propto \sigma_{\text{SDM}}^{h^*} \times  |\frac{1}{t - m^2_h + i m_h \Gamma_h}|^2 , \label{eq:ssdm} \\
\frac{d \sigma_{\text{FDM}}}{d t} & \propto  \sigma_{\text{FDM}}^{h^*} \times |\frac{1}{t-m^2_{H_1} + i m_{H_1} \Gamma_{H_1} } - \frac{1}{t-m^2_{H_2} + i m_{H_2} \Gamma_{H_2} }  |^2 \cdot \left( 2t - 8 m_\chi^2 \right) , \label{eq:sfdm}\\
\frac{d \sigma_{\text{VDM}}}{d t} & \propto  \sigma_{\text{VDM}}^{h^*} \times|\frac{1}{t-m^2_{H_1} + i m_{H_1} \Gamma_{H_1} } - \frac{1}{t-m^2_{H_2} + i m_{H_2} \Gamma_{H_2} }  |^2 \cdot \left(2+ \frac{(t-2 m^2_V)^2}{4 m_V^4} \right)   .\label{eq:svdm}
\end{align}
Detailed derivation of above relations are given in Appendix~\ref{sec:app}.  According to Eqs.~\ref{eq:ssdm}-\ref{eq:svdm}, if there is DM with mass above half of the SM Higgs boson mass (same as $H_1$ mass) while below half of the $H_2$ mass (thus $\sqrt{t} > m_{h/H_1}$), the $t$ distribution of SDM will be suppressed by the propagator at larger $t$ and that of FDM/VDM will be peaked at $m_{H_2}$. For FDM and VDM, the distributions of $t$ at tails will be also different due to the weight factors from matrix element calculation, i.e. $2t - 8 m_\chi^2$ for FDM and $2+ \frac{(t-2 m^2_V)^2}{4 m_V^4}$ for VDM. 
This point can become clear if we choose some benchmark points and show the results numerically.

There are totally four parameters of relevance in the FDM model for collider phenomenology: $g_{\chi}$, $\sin\alpha$, $m_\chi$ and $m_{H_2}$. The benchmark points are chosen to guarantee sufficient DM production rates at colliders while consistent with current Higgs precision measurements. So we take $g_{\chi}=3$, $\sin\alpha=0.3$, $m_\chi =80$ GeV and four different $m_{H_2} = \{ 200, ~300, ~400,~500 \}$ GeV, which will be denoted by FDM200, FDM300, FDM400 and FDM500, respectively. 
The partial width of $H_2 \to H_1 H_1$ is assumed to be negligible \footnote{ 
If $H_2 \to H_1 H_1$
contributes significantly to the $H_2$ decay width, the cross section of DM signal  gets smaller. 
But  $H_2$ decay width gets wider, which could improve the DM  spin discrimination 
as discussed in this paper.}
and then $H_2$ is dominantly decay into $\chi \bar{\chi}$, e.g. Br$(H_2 \to \chi \chi) > 96\%$ 
for all benchmark points. 
We note that future precision measurement of Higgs signal strength which could reduce the allowed $\sin\alpha$ can only lead to a total rescaling in production cross section in our discussion. 

The parameters for the VDM model are chosen accordingly: $\sin\alpha = 0.3$ and $m_V =80$ 
GeV.  The decay width of $H_2$ is an observable which may be determined from 
other measurements. Also we wish to keep the branching ratios of $H_2 \to V V$ the same as 
those of  $H_2 \to \chi \chi$. Therefore the $g_V$ for each benchmark point is chosen to keep 
the total decay width of $H_2$ the same with that in the FDM case. 
Table~\ref{tab:gv} provides the $g_V$ values of VDM 
benchmark points (VDM200, VDM300, VDM400 and VDM500) and the corresponding $H_2$ 
decay widths. 
\begin{table}[htb]
\begin{center}
\begin{tabular}{|c||c|c|c|c|} \hline
$m_{H_2}$ [GeV] & 200 & 300 & 400 & 500 \\ \hline
$\Gamma(H_2)$ [GeV] & 14.2 & 60.1 & 103.0 & 144.5 \\ \hline
$g_V$ & 3.53 & 3.07 & 2.37 & 1.91 \\ \hline
\end{tabular}
\caption{\label{tab:gv} The values of $g_V$  in VDM model  and its corresponding decay widths of $H_2$.  }
\end{center}
\end{table}

As for the SDM model, there are only two free parameters: $m_S$ and $\lambda_{HS}$. To coincide with the choice in FDM model, in the following study of DM spin discrimination, we take $m_S=80$ GeV and $\lambda_{HS}$ is chosen such that the number of signal events after all selections are kept the same as that of each benchmark point of the FDM model. 
However, changing the $\lambda_{HS}$ can only lead to total rescaling of the signal cross section and will not affect the kinematic variable distributions in the SDM case. 

\begin{figure}[htb]
\begin{center}
\includegraphics[width=0.48\textwidth]{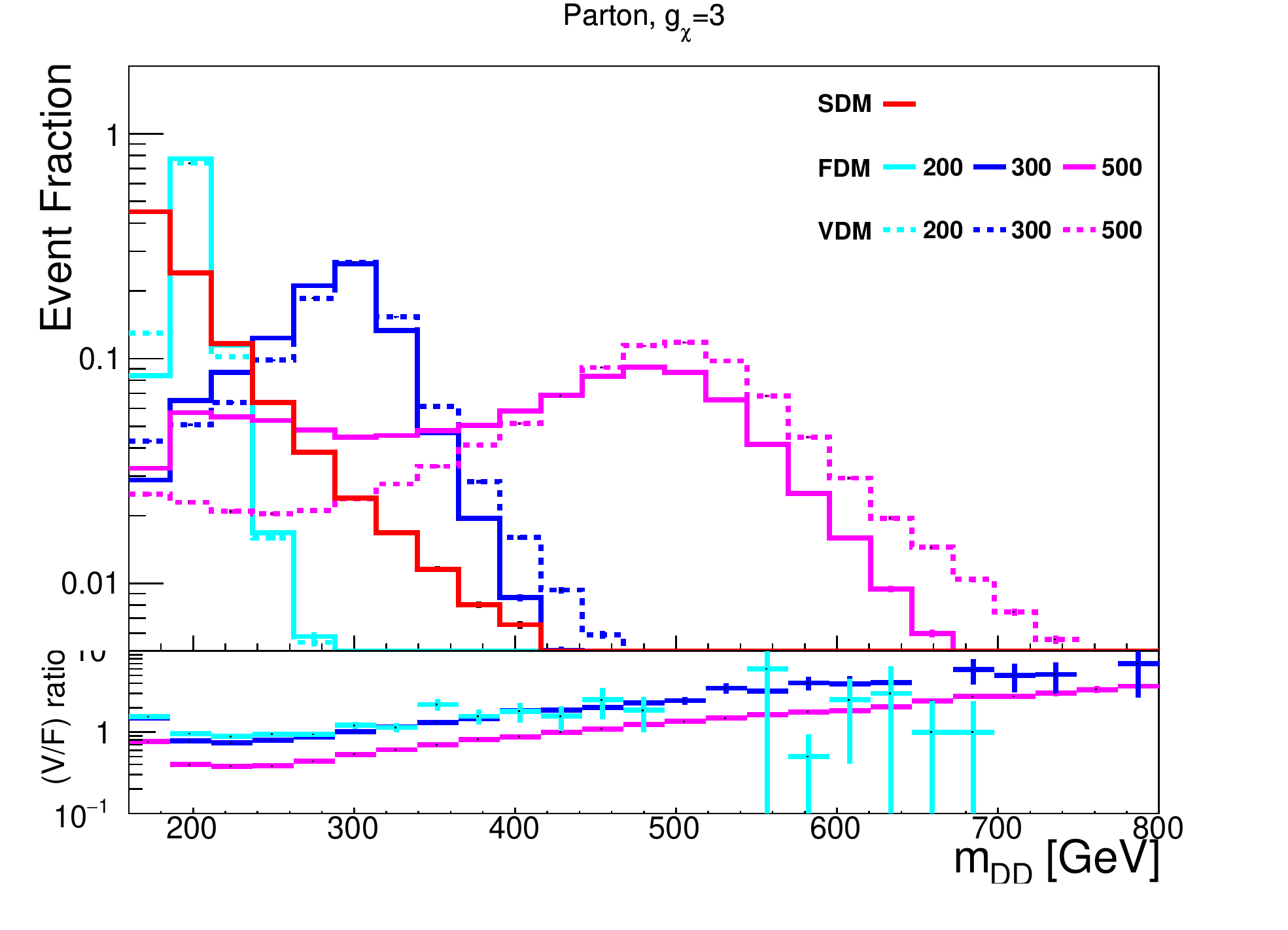}
\includegraphics[width=0.48\textwidth]{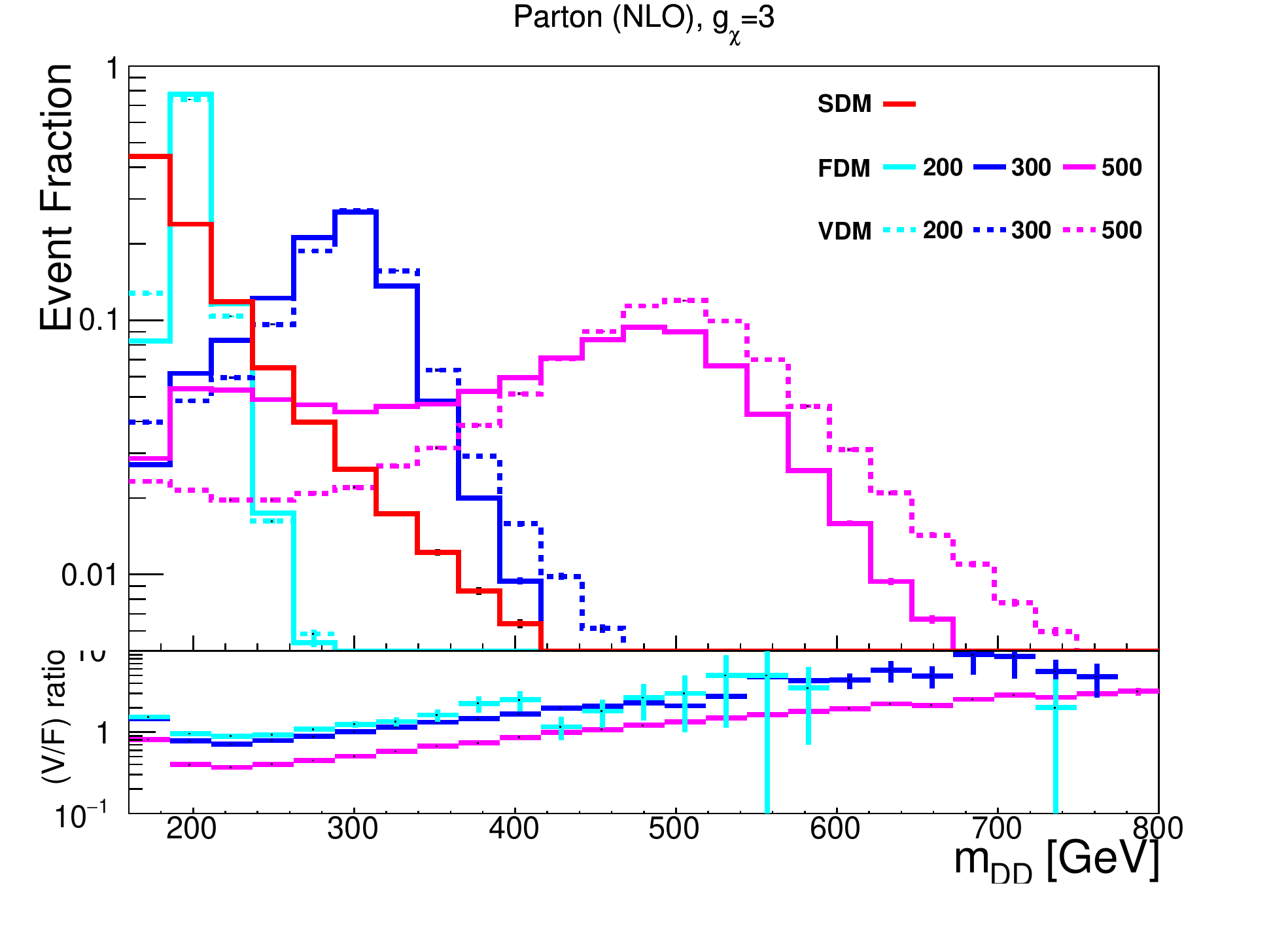}
\end{center}
\caption{\label{fig:mdd} Distributions of DM pair invariant mass in the $t\bar{t}$+DM associated production channel at 100 TeV $pp$ collider for 80 GeV DM at LO (left) and NLO (right). 
The lower plots show the ratios between the event fractions of FDM and VDM. The dashed lines correspond to the benchmark points in VDM model which have the same $H_2$ masses with the 
points in FDM model that are shown by the solid lines with the same colors. All distributions are normalized to unity. } 
\end{figure}

Based on those proposed benchmark points, we plot the distributions of $m_{DD} \equiv \sqrt{t}$ 
for the DM pair production through the $t\bar{t}$ associate channel at 100 TeV $pp$ collider in 
Fig.~\ref{fig:mdd}. 

In the left panel of the figure, we can see that for the SDM, the event fraction is largest for $m_{DD} \sim 2 m_D$ and drops quickly with increasing $m_{DD}$ due to the propagator as well as the phase space suppression. 
The $m_{DD}$ distributions for benchmark points of FDM model are peaked at $m_{DD} \sim m_{H_2}$ because of the resonant enhancement. We can also observe the interesting interference effects between two scalar mediators~\cite{Ko:2016ybp}: (1)  
the destructive interference in the region $m_{DD} > m_{H_1/ H_2}$, e.g.  the distribution of FDM200 is dropping more rapidly than SDM in the region $m_{DD} \gtrsim 200$ GeV; (2) the constructive interference in the region $m_{DD} \in [m_{H_1}, m_{H_2}]$ which leads to relatively flat event fraction in this mass region. 
The event fraction distributions of VDM benchmark points follow similar features as those of FDM, because of the same propagator structure. However, as we have calculated before, the different $t$ variable dependence in the matrix element of scalar to DMs decay will leads to distinguishable deviation in the $m_{DD} \equiv \sqrt{t}$ distributions (see Eqs.~\ref{eq:ssdm}-\ref{eq:svdm}). 
Since the weight factor of FDM  depends linearly  on $t$ while that of 
VDM is quadratic, we can 
expect that VDM will have more event fraction in the large $m_{DD}$ region, as being demonstrated 
in the lower subplot. The ratio between the event fractions of VDM and FDM is smaller than unit 
when $m_{DD} \lesssim m_{H_2}$ and greater than unit when $m_{DD} \gtrsim m_{H_2}$. 
This behavior is more visible for a benchmark point with heavy $H_2$  where the resonance 
enhancement is not that severe. 
We also stress that this argument still persist when the next-to-leading order (NLO) corrections are included. In the right panel of the same figure, we plot the $m_{DD}$ distributions with the NLO 
QCD correction. All distributions are almost unaltered.  

\begin{figure}[htb]
\begin{center}
\includegraphics[width=0.48\textwidth]{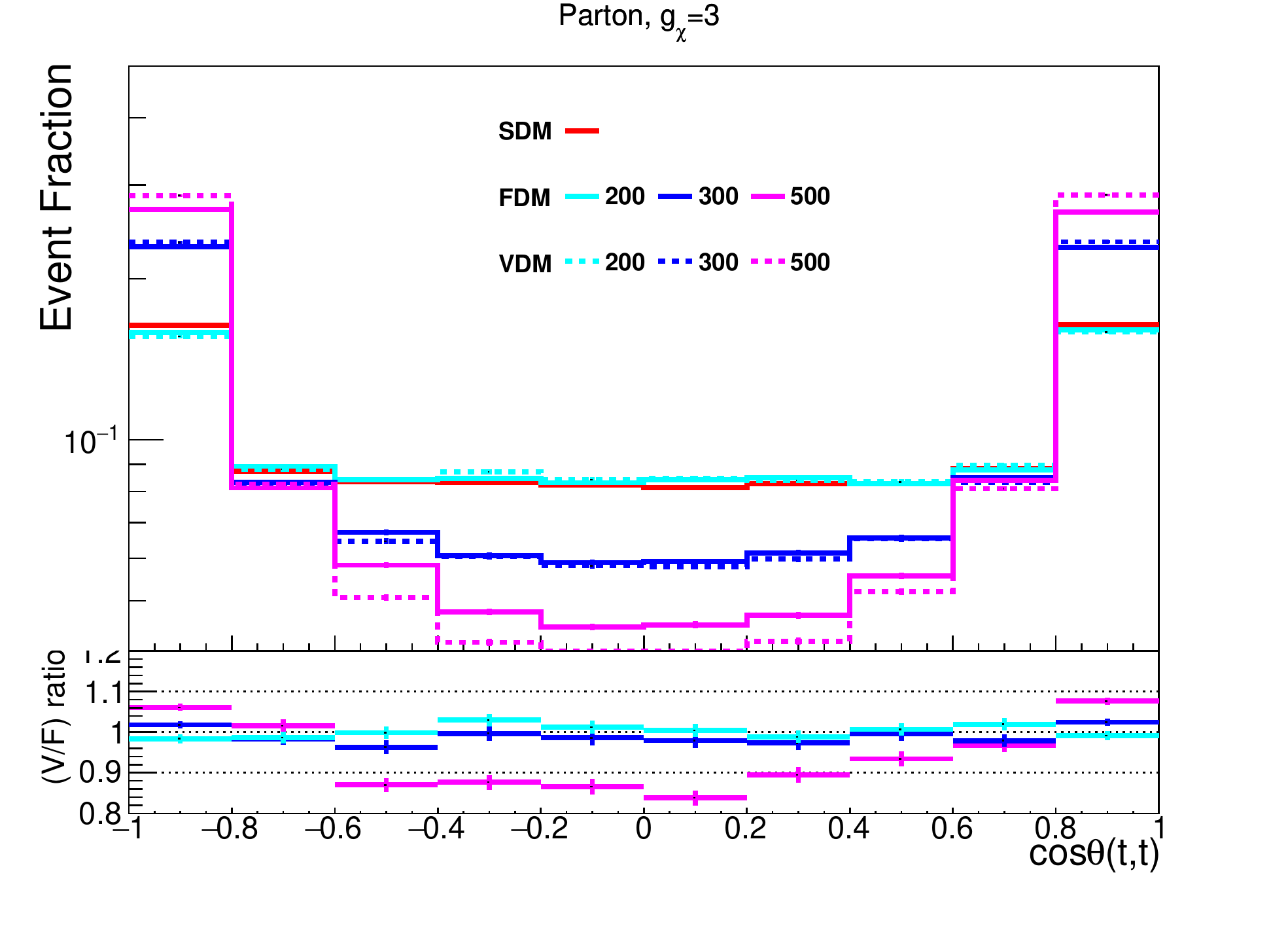} 
\includegraphics[width=0.48\textwidth]{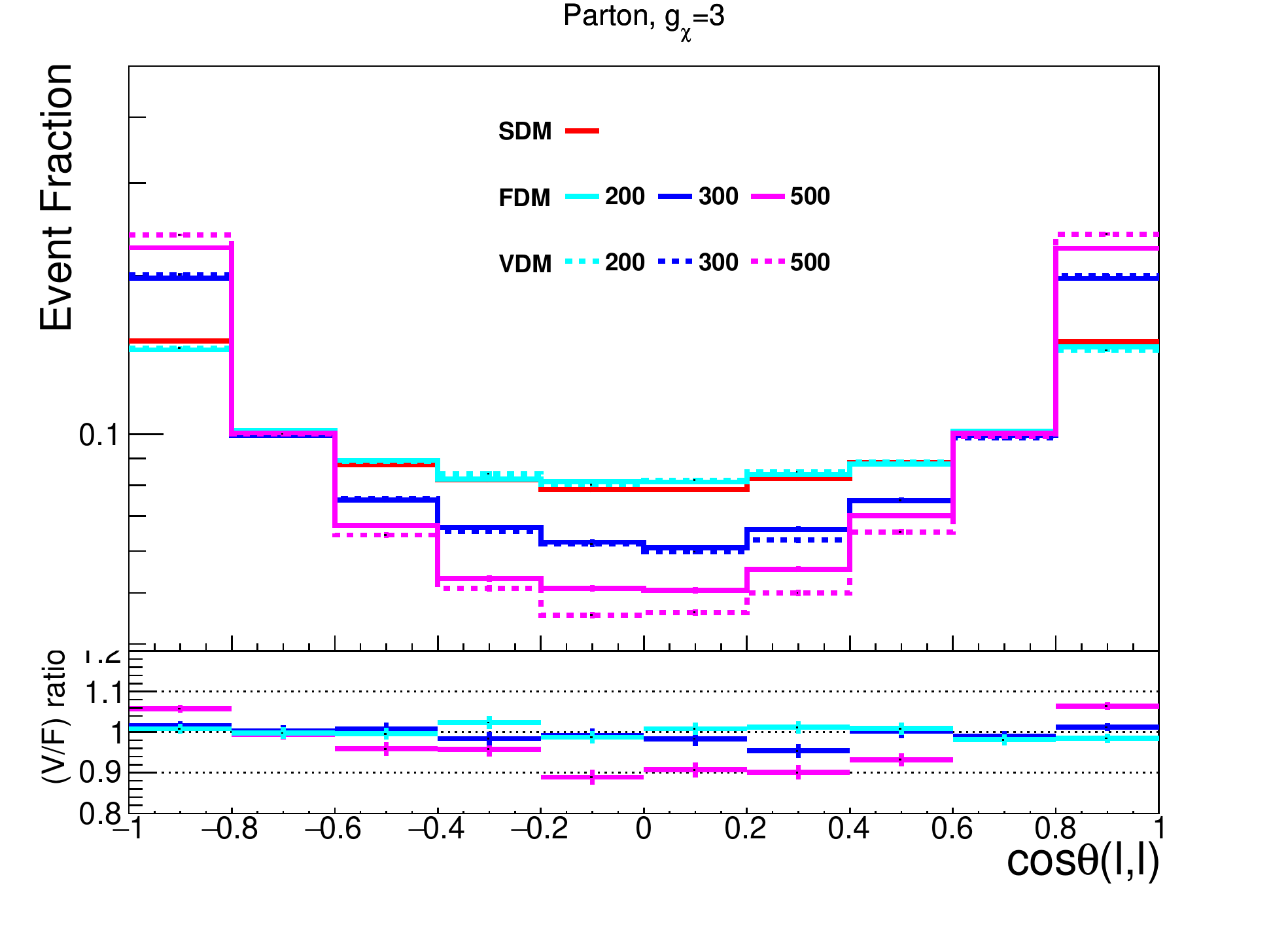}\\
\includegraphics[width=0.48\textwidth]{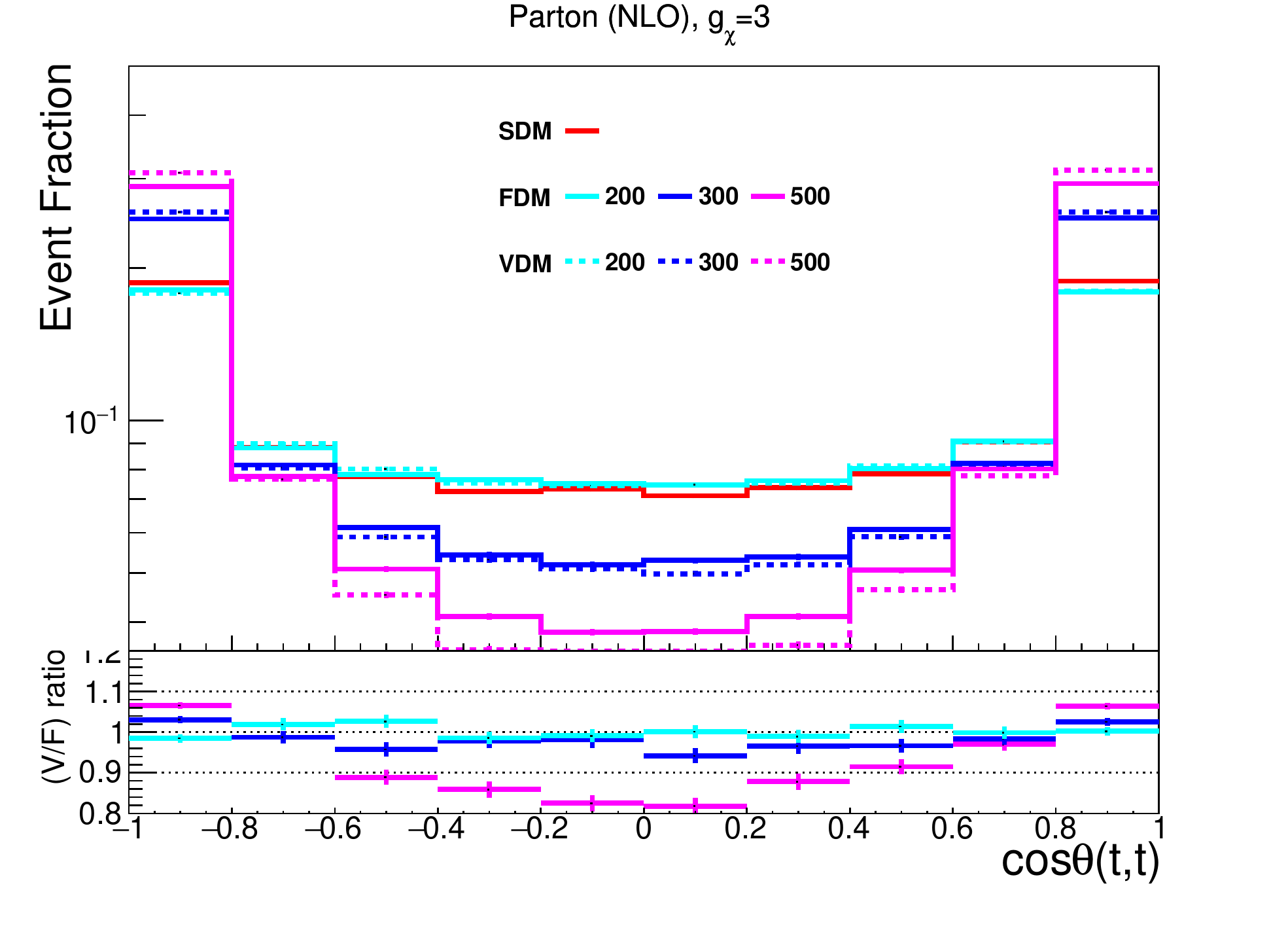} 
\includegraphics[width=0.48\textwidth]{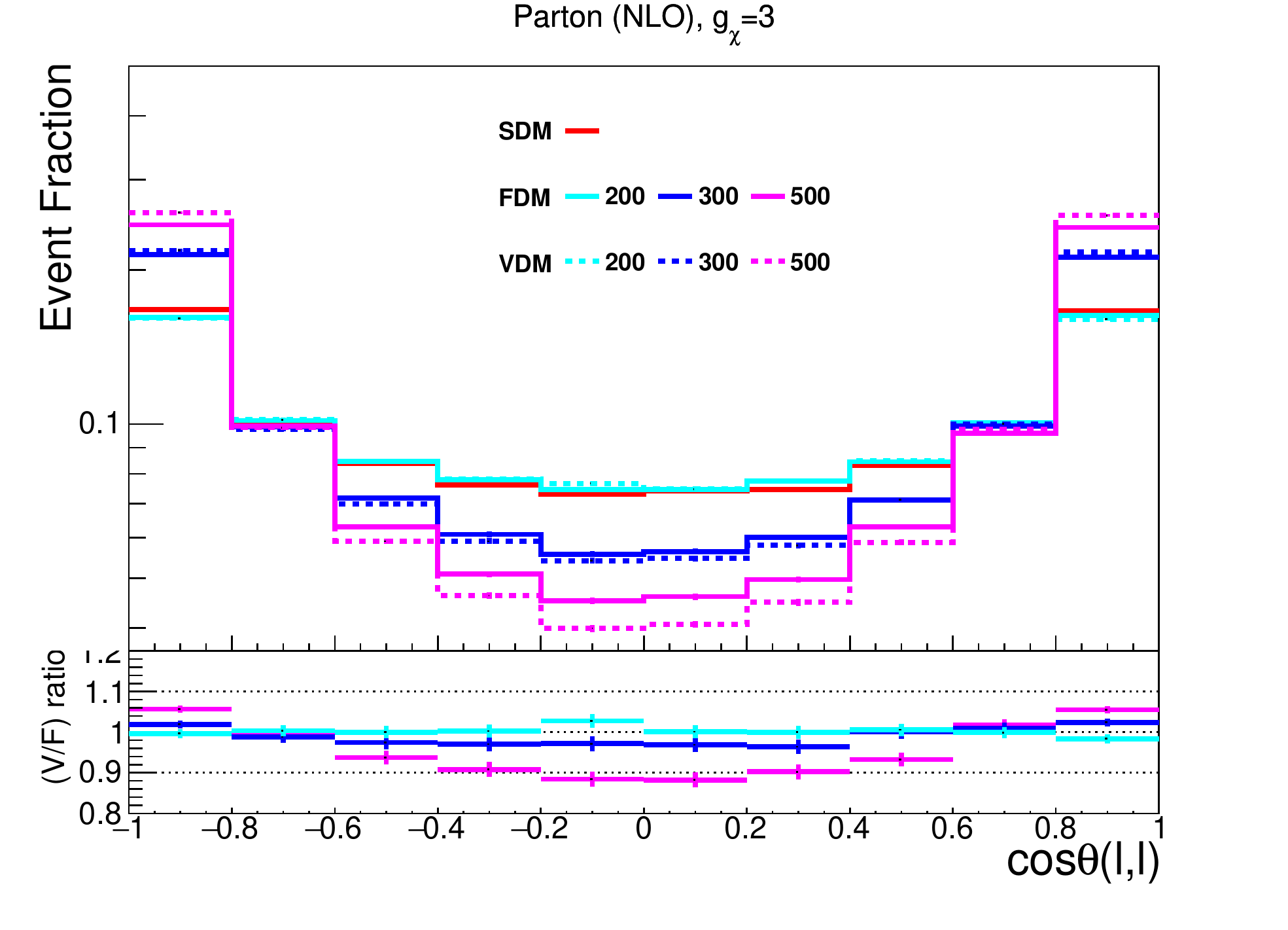}
\end{center}
\caption{\label{fig:theta} Distributions of the angular difference between top quark pair (left) and the lepton pair from top quark decay (right) at leading order (upper) and next-to-leading order (lower). }
\end{figure}

We know that the $m_{DD}$ is corresponding to the off-shell mass of a virtual scalar that is produced recoiling against two top quarks. For a given collision energy at $pp$ collider, a heavier virtual scalar would lead to less energy in the recoiling top quark pair, thus larger angular difference between the two top quarks. There are two angular variables that can be used to characterize the top quark separation: the azimuthal angle difference $\Delta \phi(t,t)$ and the polar angle difference 
$\cos\theta(t,t) \equiv \tanh (\Delta \eta(t,t)/2)$. We find they work equally well in our study so 
we simply focus on the polar angle difference throughout this work~\cite{Haisch:2016gry}. 
In the upper-left panel of Fig.~\ref{fig:theta}, the distributions of the $\cos\theta(t,t)$ for all benchmark points are presented. Comparing to the Fig.~\ref{fig:mdd}, we can find the high correlation between the $m_{DD}$ and $\cos\theta(t,t)$ distribution: (1) the SDM has quite similar $\cos\theta(t,t)$ shape with the FDM200/VDM200 since their $m_{DD}$ distributions are close; (2) for either FDM or VDM case, with the increasement of $H_2$, $m_{DD}$ is distributed toward larger value, which in turn leads to larger angular separation; (3) the difference between FDM and VDM is still appreciable in the $\cos\theta(t,t)$ distribution. 

However, we are considering the dileptonic decaying top quark pair of this channel. 
It will be impossible to reconstruct the directions of the two tops at the detector, because of multiple 
invisible particles in the final states. On the other hand, the direction of the charged lepton from the 
top quark decay is correlated to the top quark spin axis, so the angular variables of the leptons can 
be used as proxies for the top quark angles. The distributions of polar angle difference between two 
leptons ($\cos\theta(\ell,\ell)$) are given in the upper-right panel of Fig.~\ref{fig:theta}, which indeed 
look similar to the distributions of $\cos\theta(t,t)$. The smearing effect due to this indirect 
measurement makes the distinction among different scenarios slightly harder. 

Finally, we also show the distributions of both $\cos\theta(t,t)$ and $\cos\theta(\ell,\ell)$ at NLO in lower plots of Fig.~\ref{fig:theta}. Because the differences in both $\cos\theta(t,t)$ and $\cos\theta(\ell,\ell)$ mainly originate from the $m_{DD}$ distributions which is however not altered by the NLO correction, the changes in the distributions of $\cos\theta(t,t)$ and $\cos\theta(\ell,\ell)$ after considering NLO effect are found to be quite small,  even though we can observe slightly increased 
deviations among different scenarios according to the Monte Carlo simulation.

\section{Collider searches}
\label{sec:collider}

We generate the signals and SM backgrounds events at NLO level within the framework of MadGraph5\_aMC@NLO program~\cite{Alwall:2014hca,Hirschi:2015iia}. 
The UFO model files which include the NLO QCD counterterms are generated by the FeynRules~\cite{Alloul:2013bka,Degrande:2014vpa}. MadSpin
~\cite{Artoisenet:2012st} is used to generate decays of the top quark and the $W$ boson in the final 
state in order to maintain the angular information of the decay products. 
The Pythia8~\cite{Sjostrand:2007gs} is used for parton showering and hadronization. 
The final state jets are clustered using anti-k$_T$  algorithm with parameter $R=0.4$ as 
implemented in Fastjet~\cite{Cacciari:2011ma}. Finally, the detector effects are simulated by using Delphes~\cite{deFavereau:2013fsa}, where we adopt the ATLAS configuration card to mimic the smearing and reconstruction efficiency at future collider
\footnote{We expect to get similar results with CMS configuration card.}. 
The $b$-tagging efficiency~\cite{Aad:2014vea} is set to be 70\%, and the corresponding mis-tagging 
rates for the charm- and light-flavor jets are taken to be 0.15 and 0.008, respectively.

Because of suppression of the SM background and precise measurement of leptons angles, 
we only consider the dileptonic  channel in 
$t(\to b \ell \nu) \bar{t}(\to b \ell \nu) +$ DMs  production at 100 TeV $pp$ collider. 
The dominant SM background processes are associated $tW$ and $t\bar{t}$ production in dileptonic channels as well as $t\bar{t}Z(\to \nu\nu)$ production. The latter is particularly important when a hard cut on the 
$E^{\text{miss}}_T$ is applied. Moreover, we find  that the  $t\bar{t} W$ can also be subdominating, 
if the lepton in $W\to \ell \nu$ decay is not detected in the detector. 

\subsection{Search strategy}
\label{sec:ss}

Our preselection of signal events require exactly two opposite sign leptons ($e,~\mu$) and at least one $b$ jet in the final state~\footnote{We find requiring two b-jets reduces both signal and backgrounds by a factor around $1/3$. }. The leptons should have $p_T >20$ GeV and $|\eta|<2.5$, as well as be isolated: the scalar sum of transverse momenta of all particles with $p_T >0.5$ GeV that lie within a cone of radius $R=0.5$ around the $e(\mu)$ is less than 12\%(25\%) of the transverse momentum of the $e(\mu)$. The $b$-jets need to fullfil $p_T >25$ GeV and $|\eta|<2.5$ to ensure relatively high tagging efficiency. 
In the second and third rows of Table~\ref{tab:cuts}, the cross sections of backgrounds~\cite{Mangano:2016jyj} and signals before and after preselections are given. The NLO QCD corrections have been taken into account. 
For the cross section of $t\bar{t}$ process, we require at least one top quark to decay 
leptonically and the missing transverse momentum due to the neutrino in the final state to be  
larger than 100 GeV. 
Since we only required one $b$-jet in the final state, the $WWb$ with dileptonic decaying $W$ pair is also an important background for our analysis. This process is dominated by the $tW$ production with subsequent top quark decay $t\to bW$. 
We can find the preselection reduces the background cross sections by 
a factor of a few $\mathcal{O}(10)$, partly because of the branching ratio suppression. 
Signal benchmark points with different   masses are all reduced by a similar amount, i.e. 
a factor of 5, mainly  originating  from the lepton reconstruction efficiency. 

\begin{figure}[htb]
\begin{center}
\includegraphics[width=0.48\textwidth]{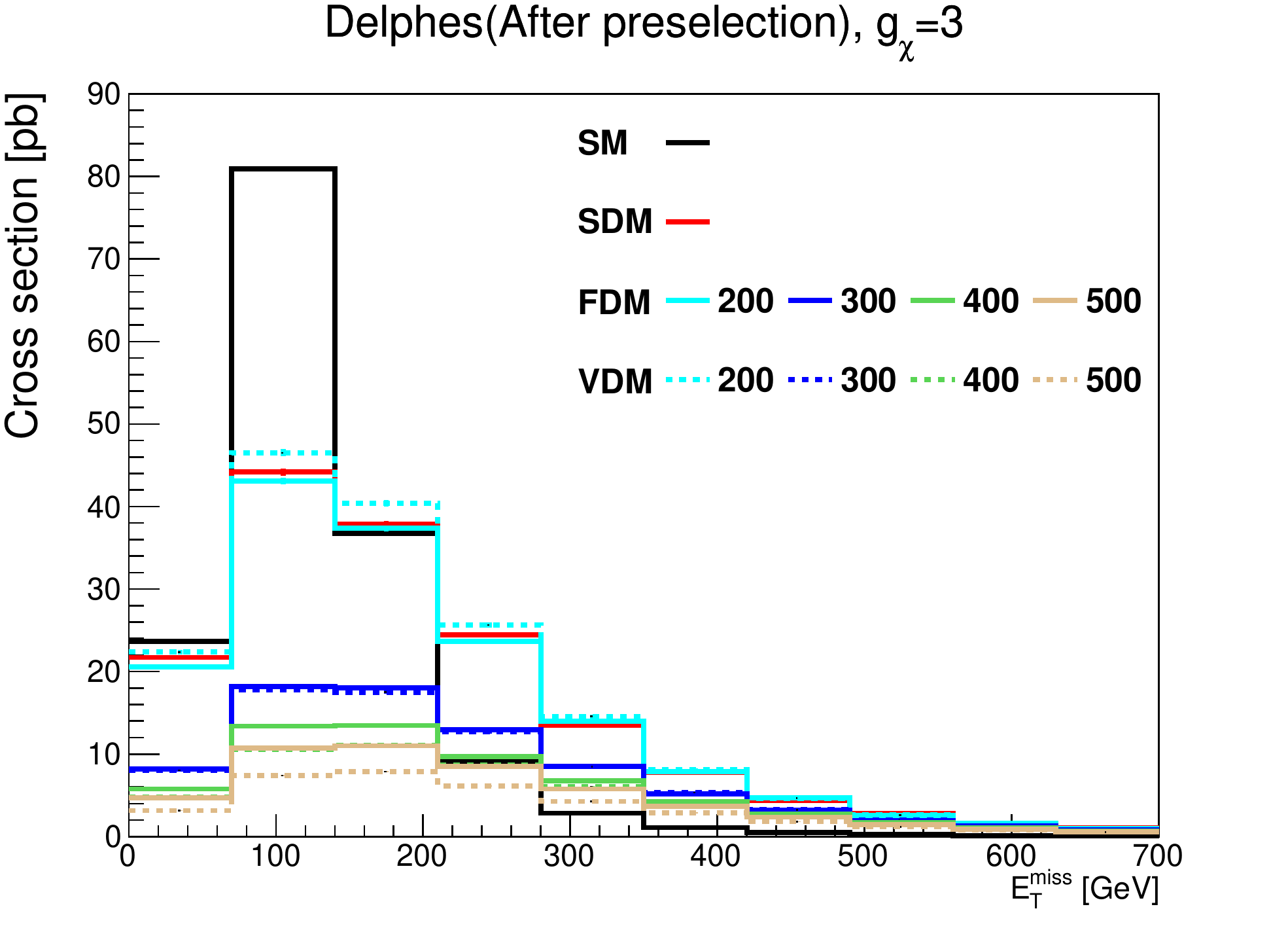}
\includegraphics[width=0.48\textwidth]{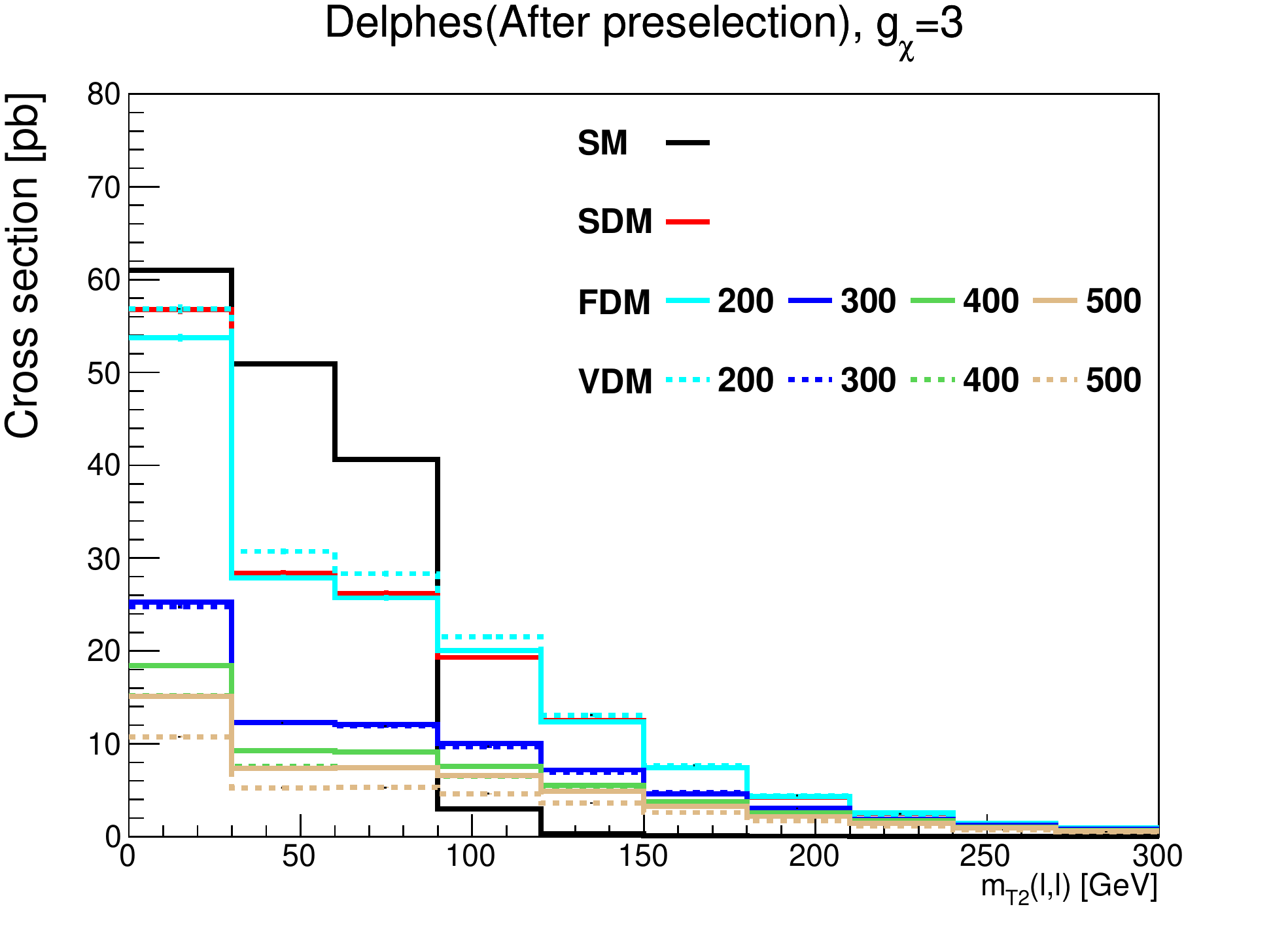}
\end{center}
\caption{\label{fig:predist} Distributions of $E^{\text{miss}}_T$ and $m_{T_2}(\ell,\ell)$ for signals and background (weighted sum) after the preselections. The distribution of background is normalized to the actual cross section and those of signals have been $2\times 10^4$ times magnified for visibility. }
\end{figure}

A few cuts on some kinematic variables are applied to further improve the signal and background discrimination. 
To reject the two leptons from $Z$ boson decay in the $t\bar{t}Z$ background, the two lepton invariant mass $m_{\ell\ell} \equiv \sqrt{(p_{\ell_1} + p_{\ell_2})^2}$ should be far from the $Z$ pole, $m_{\ell \ell} \notin [85,95]$ GeV. The cross section of $t\bar{t}Z$ after the preselection can be decreased to half after imposing this condition, while others are kept almost the same. 

For our benchmark points, the mediator of mass  around  a few $\mathcal{O}(100)$ 
GeV is produced in association with top quarks. So the signals will typically have harder spectrum in $E^{\text{miss}}_T$ than the backgrounds where the missing transverse energy is from either neutrino in top quark decay or neutrinos (and missing leptons) from vector boson decay. The distributions of $E^{\text{miss}}_T$ after preselections are ploted in the left panel of Fig.~\ref{fig:predist}. We can see that even though both signals and background distributions peak at around 100-200 GeV, the signals have much flatter tail than the background especially for benchmark points with heavy $H_2$. At this stage, we simply apply cut $E^{\text{miss}}_T > 150$ GeV. The shape information of $E^{\text{miss}}_T$ will be used later for more dedicated analysis. The efficiency of this cut can be seen in the fifth row of Table~\ref{tab:cuts}. 

Another useful and less correlated discriminating variable is the lepton pair stransverse 
mass~\cite{Lester:2014yga}, 
\begin{align}
m_{T_2} (\ell, \ell) &= \min_{\vec{p}^{D_1}_T + \vec{p}^{D_2}_T 
= \vec{p}^{\text{miss}}_T} \left( \max \left[  m_T(  \vec{p}^{\ell}_T, \vec{p}^{D_1}_T), ~ 
m_T(\vec{p}^{\ell}_T, \vec{p}^{D_2}_T)   \right]  \right)
\end{align}
where the $m_T(\vec{p}^{\ell}_T, \vec{p}^{D_1}_T) = \sqrt{2 p_T^{\ell} p_T^{D_1} 
(1-\cos \theta) }$ is the transverse mass of the $\ell_1 D_1 $ system. 
The stransverse mass has been demonstrated to be very powerful in characterizing the mass 
scalar of heavy particle which is produced in pair and subsequently decay into both visible and 
invisible particles. 
For the $t\bar{t}$ background, the two leptons in the final state come from the $W$ boson decay. 
So the $m_{T_2} (\ell, \ell)$ will drop rapidly at around $m_W$, as shown clearly in the right 
panel of Fig.~\ref{fig:predist}. We apply a relatively stringent cut on the stransverse mass variable
in order to reduce the background to a manageable level, $m_{T_2}(\ell,\ell)>150$ GeV. 
We can see from the last row of Table~\ref{tab:cuts}.  It reduces the cross sections of $t\bar{t}$ and $WWb$ backgrounds by three orders of magnitude and two orders of magnitude, respectively. 
As for $t\bar{t}V$ background and signal processes, some of the missing transverse momenta are coming from vector boson decay or DMs, the falling of the tails for which is much flatter  
than that of $t\bar{t}$ backgrounds. 
The $t\bar{t} W$ and $t\bar{t}Z$ events are reduced by factors of thirty and ten, respectively. As a consequence, the cross section of $t\bar{t}Z$ background becomes comparable to that of $t\bar{t}$ events after considering the $m_{T_2}(\ell,\ell)$ requirement.
Due to the heaviness of the mediator in signal process, this cut only reduce the signals by factors around four.   

\begin{table}[ht!]
 \begin{center}
 \scalebox{0.9}{
  \begin{tabular}{|c||c|c|c|c||c|c|c|c|c|} \hline
     & $\bar{t} t_\ell^{(E^{\text{miss}}_T >100~\text{GeV})}$ & $\bar{t} tW$ & $\bar{t} t Z$ & $W_\ell W_\ell b$ & FDM200 & FDM300 & FDM400 & FDM500 \\
    Cross section & 1316.5 pb & 20.5 pb   & 64.2 pb  & 128.4 pb &34.2 fb & 18.7 fb   & 14.8 fb & 12.5 fb  \\ \hline
    Presections & 63.76 pb & 351.8 fb & 1.9 pb & 25.4 pb  & 7.86 fb & 3.99 fb  & 3.05 fb & 2.55 fb \\ \hline
    $m_{\ell \ell} \notin [85,95]$ GeV & 59.8 pb & 330.4 fb  & 1.05 pb  & 23.9 pb &  7.47 fb& 3.82 fb & 2.92 fb  & 2.44 fb  \\
    $E^{\text{miss}}_T>150$ GeV & 17.76 pb & 69.61 fb &  261.14 fb & 3.5 pb & 4.17 fb & 2.44 fb & 1.93 fb & 1.63 fb \\
    $m_{T_2}(\ell,\ell)>150$ GeV & 23.83 fb & 1.92 fb & 32.1 fb & 38.0 fb & 0.87 fb &  0.62 fb &  0.54 fb & 0.47 fb \\ \hline
  \end{tabular}
  }
 \end{center}
  \caption{\label{tab:cuts} Cut flows for the SM background and FDM processes. 
  The cross sections of signals in the second row include the branching ratios of leptonically decaying tops.  Here  $\bar{t} t_\ell$ means that at least one (anti)top decays leptonically, $W_\ell$ corresponds to leptonic decay of W boson. }
\end{table}

In Table~\ref{tab:cuts}, the background cross sections are still around two order of magnitude larger than signal processes. With a signal significance estimator (ignoring the systematic uncertainty)
\begin{align}
\mathcal{S}=\sqrt{-2 \left( n_b \log \frac{n_s+n_b}{n_b} - n_s \right)},
\end{align}
we find that benchmark points FDM200, FDM300, FDM400 and FDM500 will be excluded at 2$\sigma$ level with the integrated luminosity of 509 fb$^{-1}$, 1001 fb$^{-1}$, 1319 fb$^{-1}$ and 1741 fb$^{-1}$, respectively. 

\begin{figure}[htb]
\begin{center}
\includegraphics[width=0.6\textwidth]{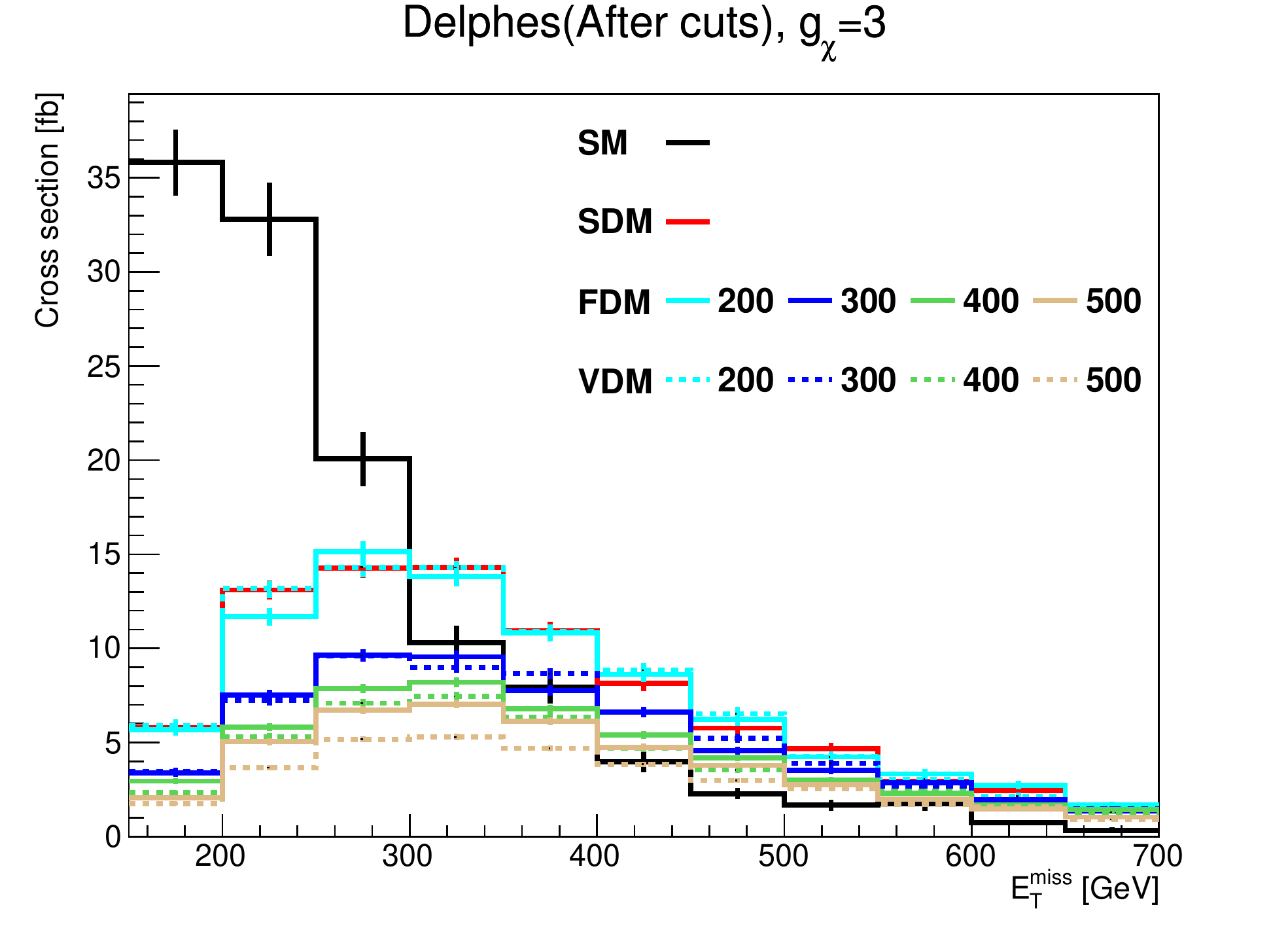}
\end{center}
\caption{\label{fig:aftermet} Distributions of $E^{\text{miss}}_T$ for signal and SM background events after applying all selection requirements. The distribution for background events is normalized to the actual cross section and those for signals have been 100 times magnified for visibility.}
\end{figure}

New, the discovery prospects are evaluated by using the shape information of $E^{\text{miss}}_T$ distributions.
Figure~\ref{fig:aftermet} shows the distributions of $E^{\text{miss}}_T$ for signals and background after applying all selection cuts. Due to the relatively high energy scale of signal processes, the event fraction of signals decrease much slower than the background with increasing $E^{\text{miss}}_T$. 
In order to quantify the difference between the signal and background in terms of the $E^{\text{miss}}_T$ distributions including the information of total normalization, we adopt the binned log-likelihood analysis~\cite{Buckley:2015ctj}.  

We first consider the $E^{\text{miss}}_T$ distribution of total background (weighted sum among all processes) as null hypothesis ($\mathcal{H}_0$) and that of background plus one of the benchmark points as test hypothesis. Due to the limited number of total events after all selections, the $E^{\text{miss}}_T$ distributions are divided into 11 bins within the range of [150, 700] GeV. In each 
bin, the probability that the $i$-th bin with the expected value of $t_i$ has $n_i$ observed events 
obeys the Poisson distribution, i.e. $({t_i^{n_i} e^{-t_i} })/ {n_i!}$. So we can determine the probability 
of the full distribution by multiplying the probability of each bin, giving the binned likelihood
\begin{align}
\mathcal{L}(\text{data}| \mathcal{H}_\alpha) = \prod_{i} \frac{t_i^{n_i} e^{-t_i} }{n_i!}. 
\end{align}
Here $i$ runs over 11 bins and $\mathcal{H}_\alpha$ corresponds different hypothesis. 
Then, we can define the test statistic $\mathcal{Q}$ as the log likelihood ratio between a given hypothesis and the null hypothesis
\begin{align}
\mathcal{Q} = -2 \log \left( \frac{\mathcal{L}(\text{data}| \mathcal{H}_\alpha)}{\mathcal{L}(\text{data}| \mathcal{H}_0 )}  \right).
\end{align}
Finally, we use the distributions of $E^{\text{miss}}_T$ in hypothesis $\mathcal{H}_0$ and 
$\mathcal{H}_{\alpha, ~\alpha>0}$ to generate two sets of pseudodata. 
Each set of pseudodata will give a distribution of the test statistics. 
Using those two distributions of $\mathcal{Q}$, we can calculate the $p$-value of the test hypothesis ($\mathcal{H}_{\alpha, ~\alpha>0}$) by assuming that the actual observation is at the center of $\mathcal{Q}$ distribution under null hypothesis.

\begin{figure}[htb]
\begin{center}
\includegraphics[width=0.6\textwidth]{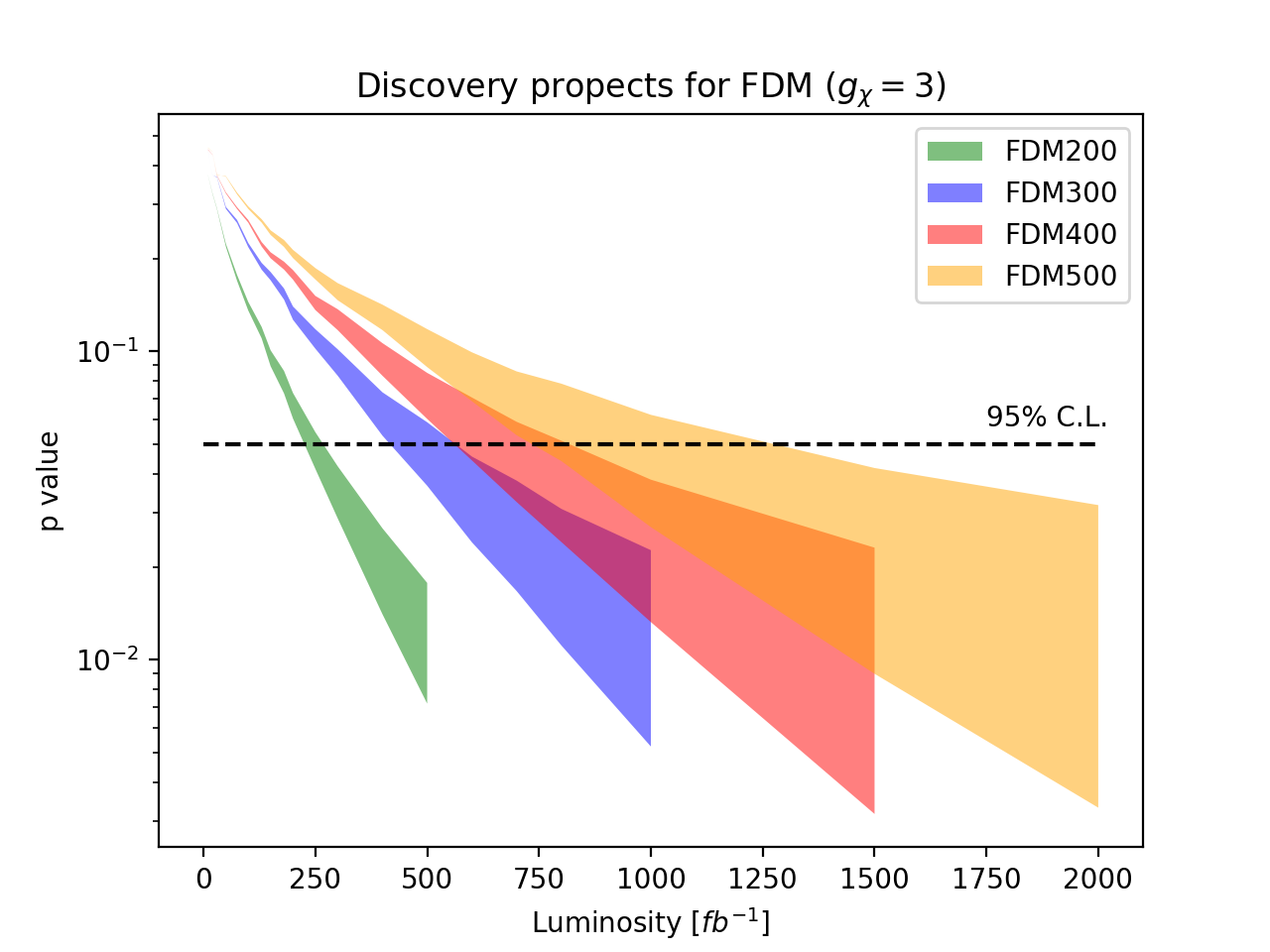}
\end{center}
\caption{\label{fig:f3disc} Discovery prospects of benchmark points in FDM model at 100 TeV $pp$ collider. The widths of bands are showing the sensitivities without systematic uncertainty (lower boundary) and assuming systematic uncertainty of 1\% (upper boundary). The $p$-value correspond to 95\% C.L. probing is indicated by the horizontal dashed line.}
\end{figure}

The $p$-values for those FDM benchmark points are shown in Fig.~\ref{fig:f3disc} with varying the integrated luminosity, where the 95\% exclusion (probing) limit is also indicated by the horizontal dashed line. 
The widths of bands are showing the sensitivities without systematic uncertainty (lower boundary) and assuming systematic uncertainty of 1\% (upper boundary)~\footnote{This optimistic estimation of systematic uncertainty has been adopted in a few studies at future collider by phenomenological group~\cite{Low:2014cba,Khoze:2015sra} and by experimental group~\cite{ATL-PHYS-PUB-2014-007,Mangano:2017tke}. }. 
By considering the shape of the $E^{\text{miss}}_T$ distributions, the required integrated luminosities for 2$\sigma$ sensitivity are roughly reduced by half for all benchmark points (250 fb$^{-1}$, 500 fb$^{-1}$, 750 fb$^{-1}$ and 1000 fb$^{-1}$ for FDM200, FDM300, FDM400 and FDM500 with systematic uncertainty less than $\sim 1\%$, respectively). 
Our benchmark points will be tested in an early stage at the future $pp$ collider. 

\subsection{Discrimination prospects}

Once an excess in dilepton + $E^{\text{miss}}_T$ events is observed, it will be important to identify the underlying new physics. 
This subsection is devoted to distinguish the benchmark points with different DM spins as proposed in Sec.~\ref{sec:model}. 

\begin{figure}[htb]
\begin{center}
\includegraphics[width=0.48\textwidth]{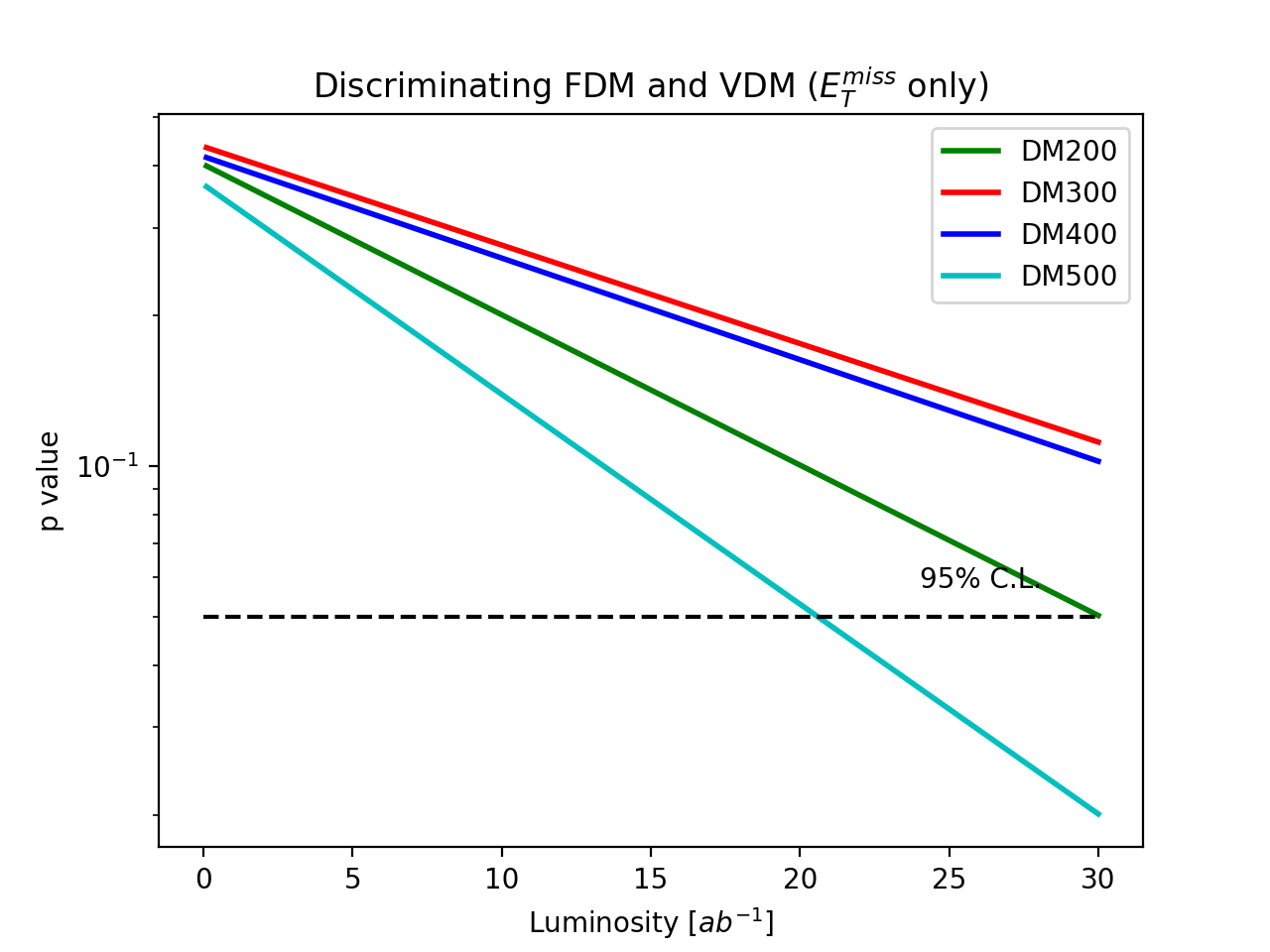}
\includegraphics[width=0.48\textwidth]{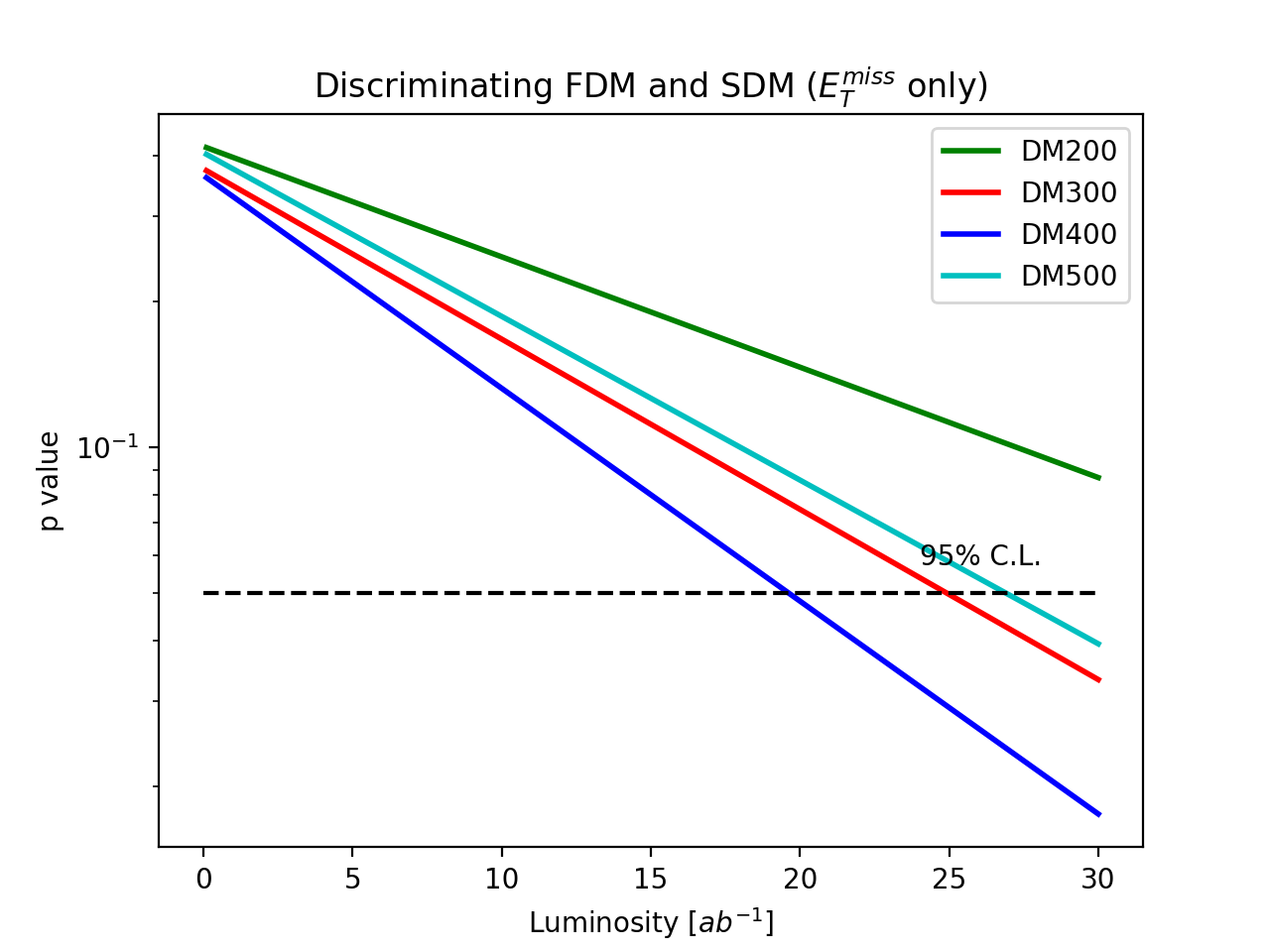}
\end{center}
\caption{\label{fig:fvs3diff} Spin discrimination prospects between FDM and VDM in the left panel; between FDM and SDM in the right panel. Only the shape information of $E^{\text{miss}}_T$ distributions are used. And the systematic uncertainty is ignored.   }
\end{figure}

As seen in Fig.~\ref{fig:aftermet}, the $E^{\text{miss}}_T$ distribution of SDM is similar to that of FDM/VDM with $H_2=200$ GeV, but it is quite different from those with heavier $H_2$. Given $H_2$ mass, the VDM has harder $E^{\text{miss}}_T$ spectrum than FDM due to the same reason as explained for $m_{DD}$ distribution in previous section (see discussions on Fig.~\ref{fig:mdd}).  
Moreover, the signal rates are also different between benchmark points of FDM and VDM, especially when the $H_2$ is heavy. 
This motivates us to study the spin discrimination by using the binned log-likelihood test again. 
But here, for each $H_2$ mass, the null hypothesis is the SM background plus a FDM benchmark 
point and the test hypothesis is the SM background plus the corresponding benchmark point of 
VDM or SDM. We note that the benchmark point of SDM model has the same number of events 
after all selections with that of FDM model. 

The $p$-values for spin discrimination with varying integrated luminosity are plotted in Fig.~\ref{fig:fvs3diff}. The future 100 TeV $pp$ collider will be able to accumulate approximately 
$30$ ab$^{-1}$ data~\cite{Zimmermann:2016puu}. 
It will be possible to distinguish FDM and VDM when the mediator 
($H_2$) mass is either light ($m_{H_2} \lesssim 200$ GeV) or heavy ($m_{H_2} \gtrsim 500$ GeV), since the production rate is large in the former case and difference in  $E^{\text{miss}}_T$ distribution 
is large in the latter case. The future $pp$ collider is not able to resolve the DM spin for $m_{H_2} \sim [300,400]$ GeV. 
For FDM and SDM, it will be possible to distinguish for benchmark points with relatively large $m_{H_2}$. 
As we have already seen from Fig.~\ref{fig:aftermet},  the $E^{\text{miss}}_T$ shapes of FDM 
and SDM become too similar for $m_{H_2} \sim 200$ GeV.  

\begin{figure}[htb]
\begin{center}
\includegraphics[width=0.24\textwidth]{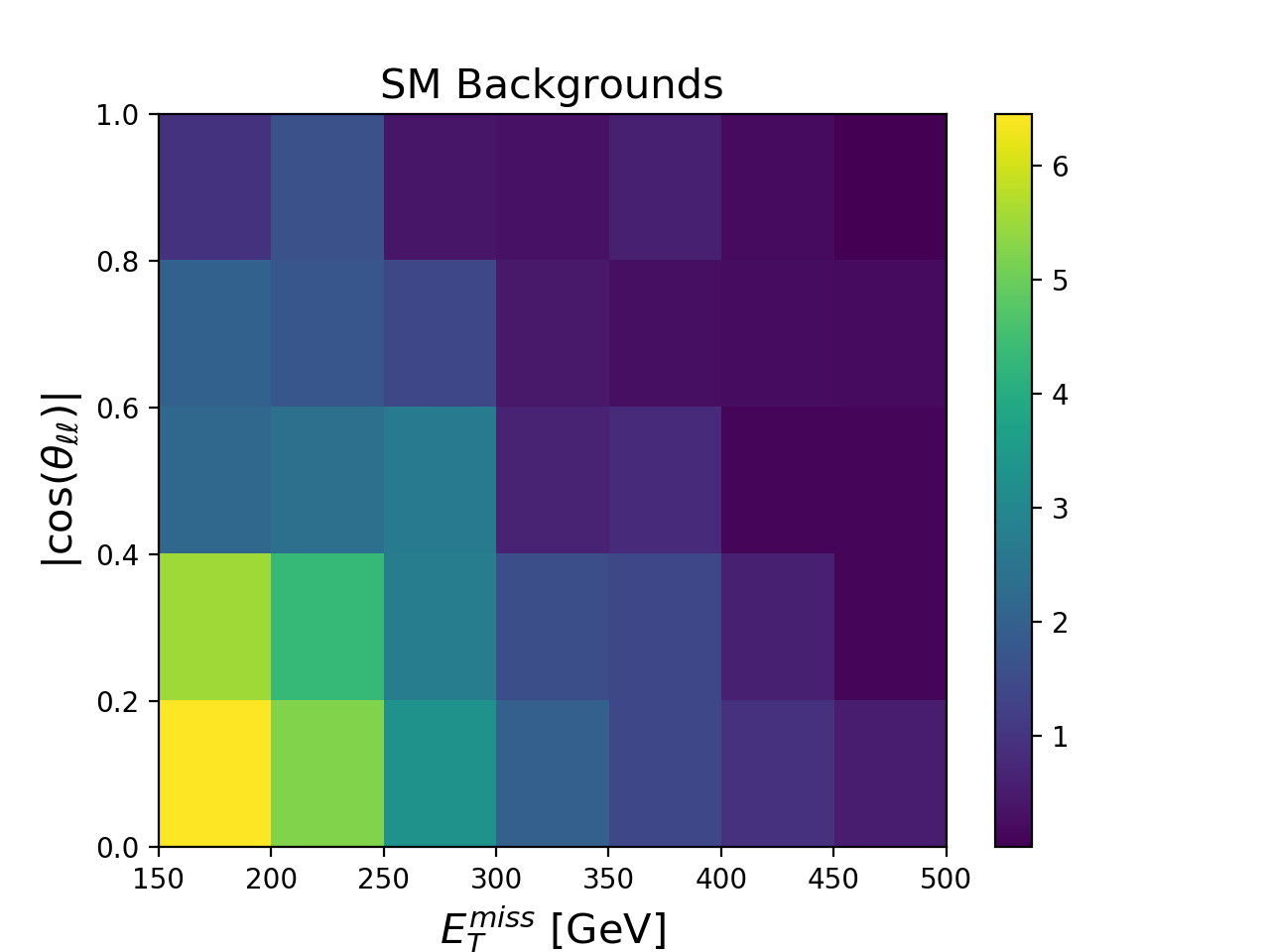}
\includegraphics[width=0.24\textwidth]{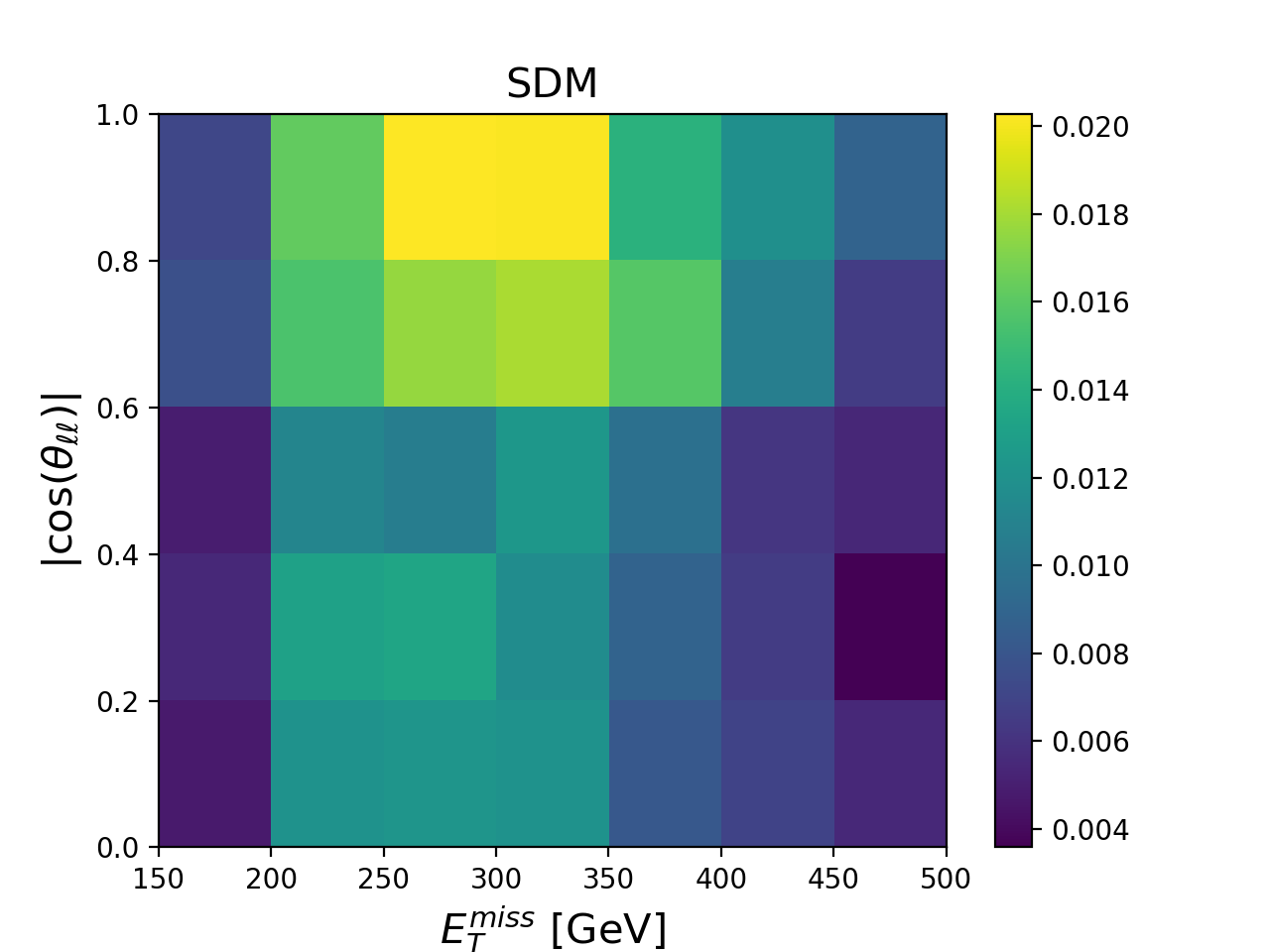}
\includegraphics[width=0.24\textwidth]{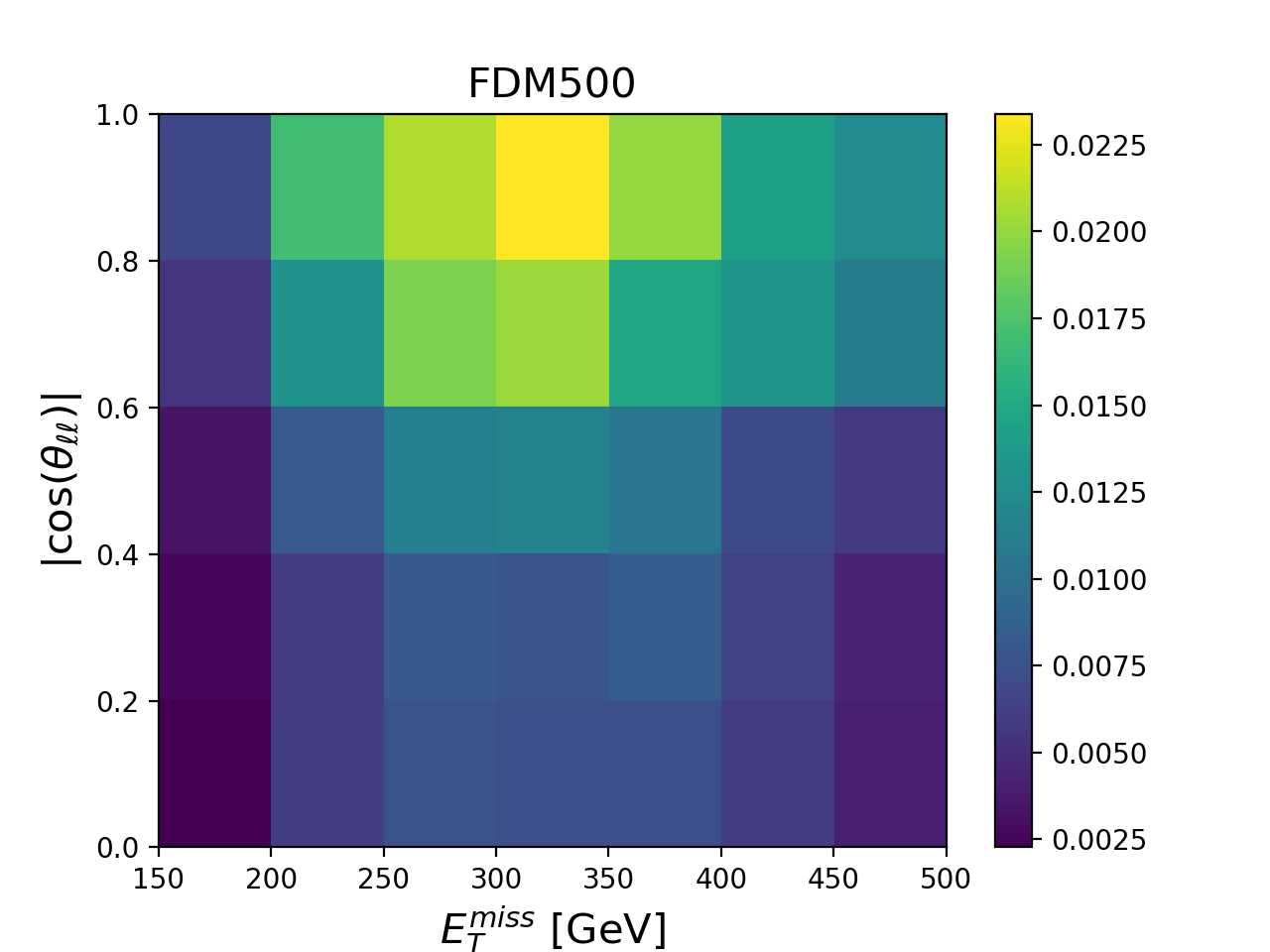}
\includegraphics[width=0.24\textwidth]{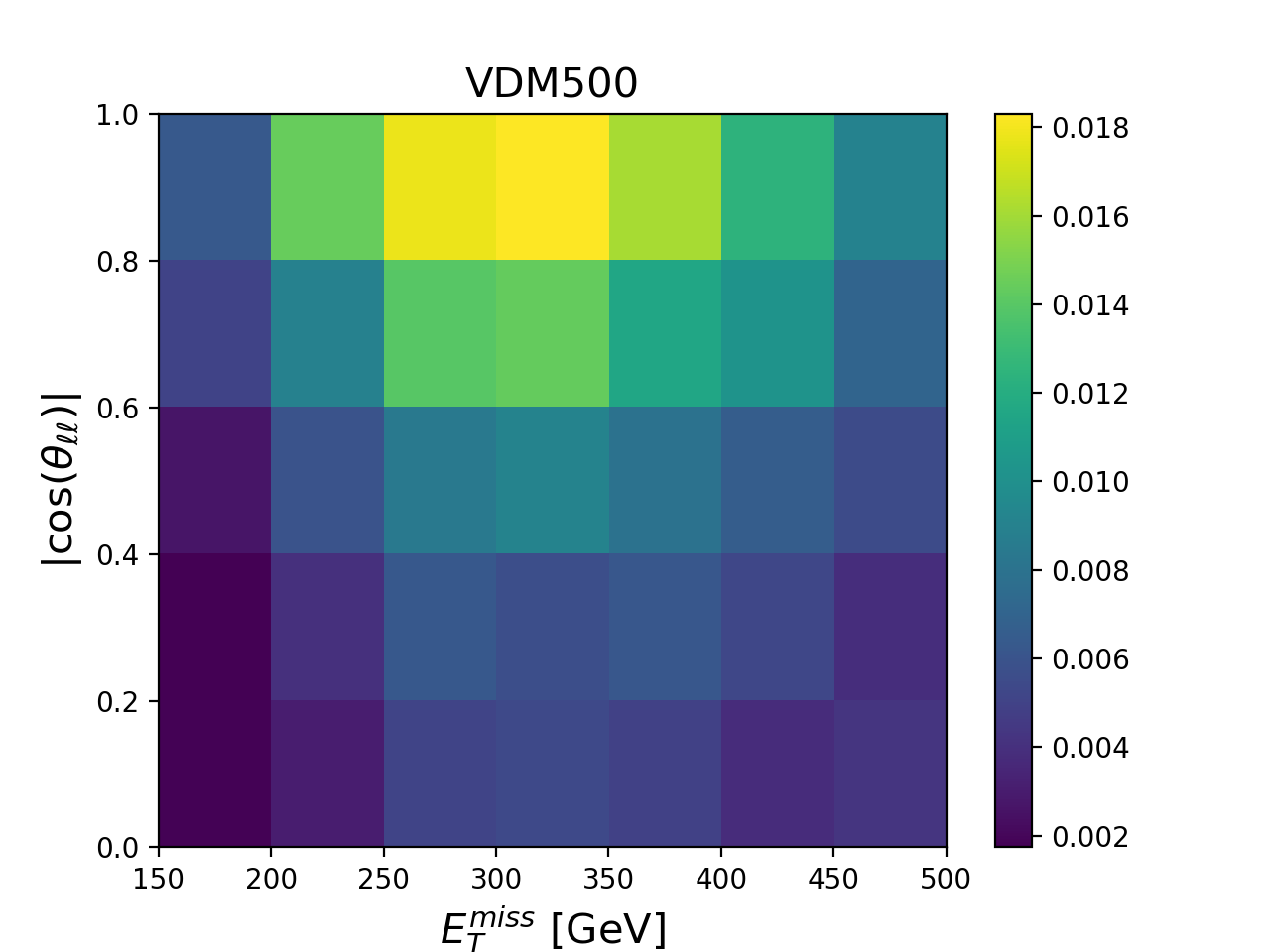}
\end{center}
\caption{\label{fig:2d} Two dimensional distributions in $|\cos(\theta_{\ell \ell})|$ and $E^{\text{miss}}_T$ plane after all selection requirements for an weighted sum of all SM backgrounds and SDM, FDM500 and VDM500, respectively. Color code is indicating the actual cross section [fb] in each bin. }
\end{figure}

To improve the spin discrimination power, the polar angle between two leptons $\cos( \theta_{\ell\ell})$ is additionally considered with $E^{\text{miss}}_T$. We perform the binned log-likelihood test on the two dimensional distribution of these two variables. Figure~\ref{fig:2d} gives the two dimensional distribution of $E^{\text{miss}}_T$ and $|\cos(\theta_{\ell \ell})|$ for {a weighted sum of SM backgrounds and} benchmark points SDM, FDM500 and VDM500. Note that the distribution of $\cos(\theta_{\ell \ell})$ is an even function, so its absolute value has been used in the histograms with five bins to maintain sufficient statistics. 

\begin{figure}[htb]
\begin{center}
\includegraphics[width=0.48\textwidth]{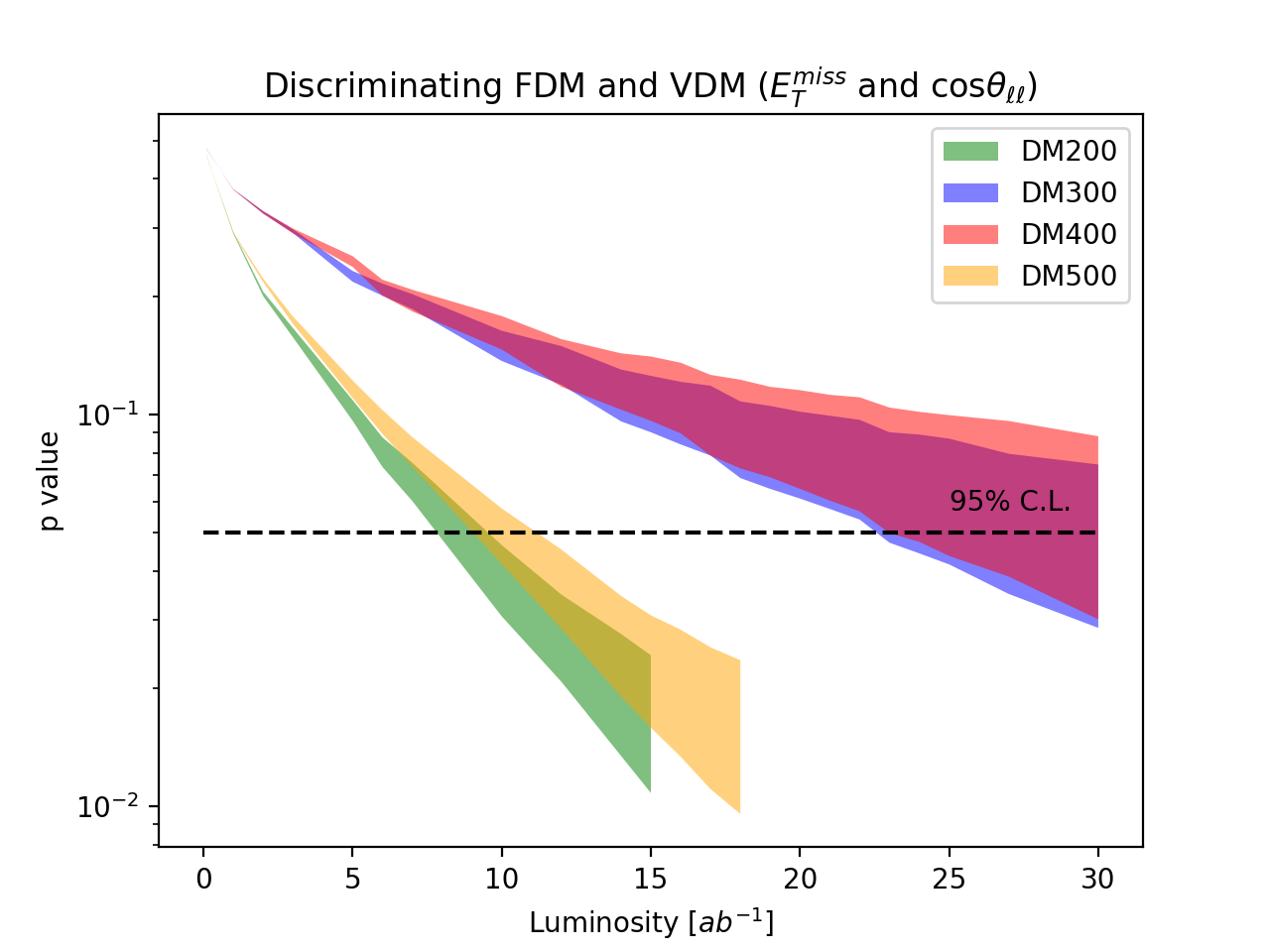}
\includegraphics[width=0.48\textwidth]{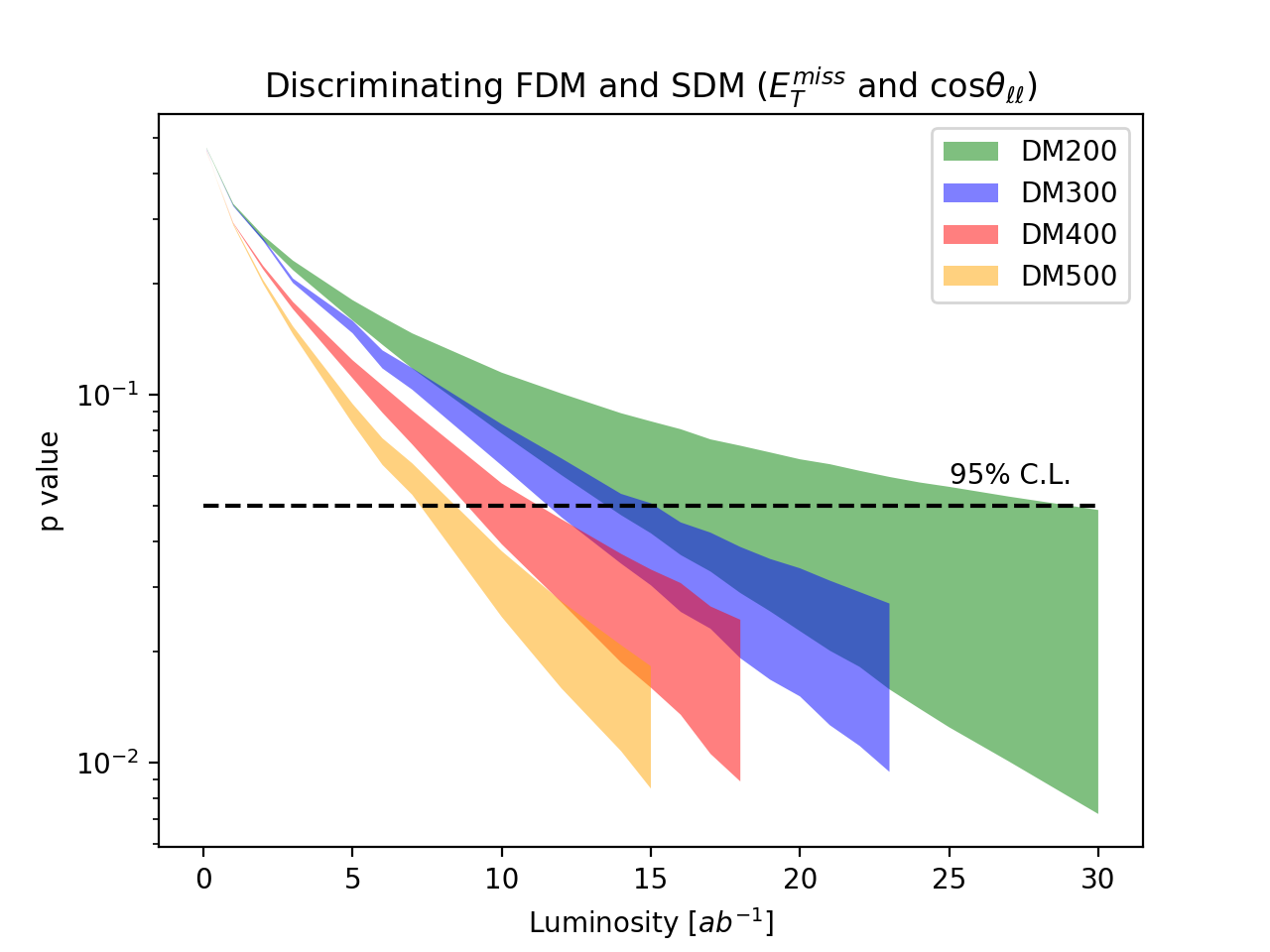}
\end{center}
\caption{\label{fig:fvs32d} Spin discrimination prospects between FDM and VDM in the left panel; between FDM and SDM in the right panel. The two dimension distributions in the $\cos(\theta_{\ell \ell})$ and $E^{\text{miss}}_T$ plane are used. Bands are plotted in the same way as explained in the caption of Fig.~\ref{fig:f3disc}, but here the systematic uncertainty is assumed to be slightly smaller, i.e. 0.5\%.  }
\end{figure}

The binned likelihood are defined with two dimensional bins
\begin{align}
\mathcal{L}(\text{data}| \mathcal{H}_\alpha) = \prod_{i,j} \frac{t_{ij}^{n_{ij}} e^{-t_{ij}} }{n_{ij}!}, 
\end{align}
where the indexes $i$ and $j$ run over the bins of $E^{\text{miss}}_T$ and $|\cos(\theta_{\ell \ell})|$ respectively. The expected $p$-values  with respect to the integrated luminosity are plotted in Fig.~\ref{fig:fvs32d}. 
Comparing to the discrimination with only $E^{\text{miss}}_T$ distribution, we can find that with the additional information from the angular separation of dilepton, the required integrated luminosities for 95\% C.L. probing are reduced by more than half for those benchmark points. 
The spin discrimination between FDM and VDM is possible at future $pp$ collider for mediator mass either $\sim 200$ GeV or $\sim 500$ GeV, where we have assumed the systematic uncertainty can be controlled at $\sim 0.5\%$ level. And the spin discrimination between FDM and SDM is even better, which can be accomplished with integrated luminosity below $\sim 30$ ab$^{-1}$ for all mediator masses given a 0.5\% systematic uncertainty.

\subsection{Varying the couplings}

We here repeat the study of the prospects for discovery and spin discrimination for $g_\chi=1$ in the FDM model, instead of $g_\chi=3$ in the previous subsections.
The benchmark points in VDM model are chosen such that the decay widths of $H_2$ are kept the same as the ones in the FDM model assuming negligible $H_2 \to H_1 H_1$ partial decay width.
In the case of the SDM models, the benchmark points are chosen such that the signal yields after the all selection cuts are kept the same with each of benchmark points in the FDM model by taking appropriate $\lambda_{HS}$ values.

\begin{table}[htb]
\begin{center}
\begin{tabular}{|c||c|c|} \hline
\multirow{ 2}{*}{$m_{H_2}$ [GeV]} & \multicolumn{2}{c|}{$g_\chi=1$}   \\ \cline{2-3}
  & FDM & VDM \\ \hline \hline
200 & 35.7 fb & 36.0 fb  \\ 
300 & 17.0 fb & 16.9 fb  \\ 
400 & 10.8 fb & 10.5 fb   \\
500 & 7.41 fb & 6.95  fb  \\ \hline
\end{tabular}
\caption{\label{tab:g110} The production cross sections for benchmark points with $g_\chi=1$. The branching ratio of leptonic top quark decay ($t\to b e \nu / b \mu \nu$) has been taken into account.  }
\end{center}
\end{table}

The signal production cross section for benchmark points of FDM and VDM models are given in Table~\ref{tab:g110}.  Comparing with Table~II for $g_\chi =3$,  we find that  the FDM signal cross sections are almost irrelevant to the coupling $g_\chi$ for relatively light $H_2$ 
($m_{H_2} \sim 200$ GeV).   This is because  
the signal is dominated by the on-shell $H_2$ production which mostly decays into DM pair. 
It should be noted that VDM will typically have lager 
cross section when $H_2$ is lighter and have smaller cross section when the $H_2$ is heavier
than the FDM. 

\begin{figure}[htb]
\begin{center}
\includegraphics[width=0.48\textwidth]{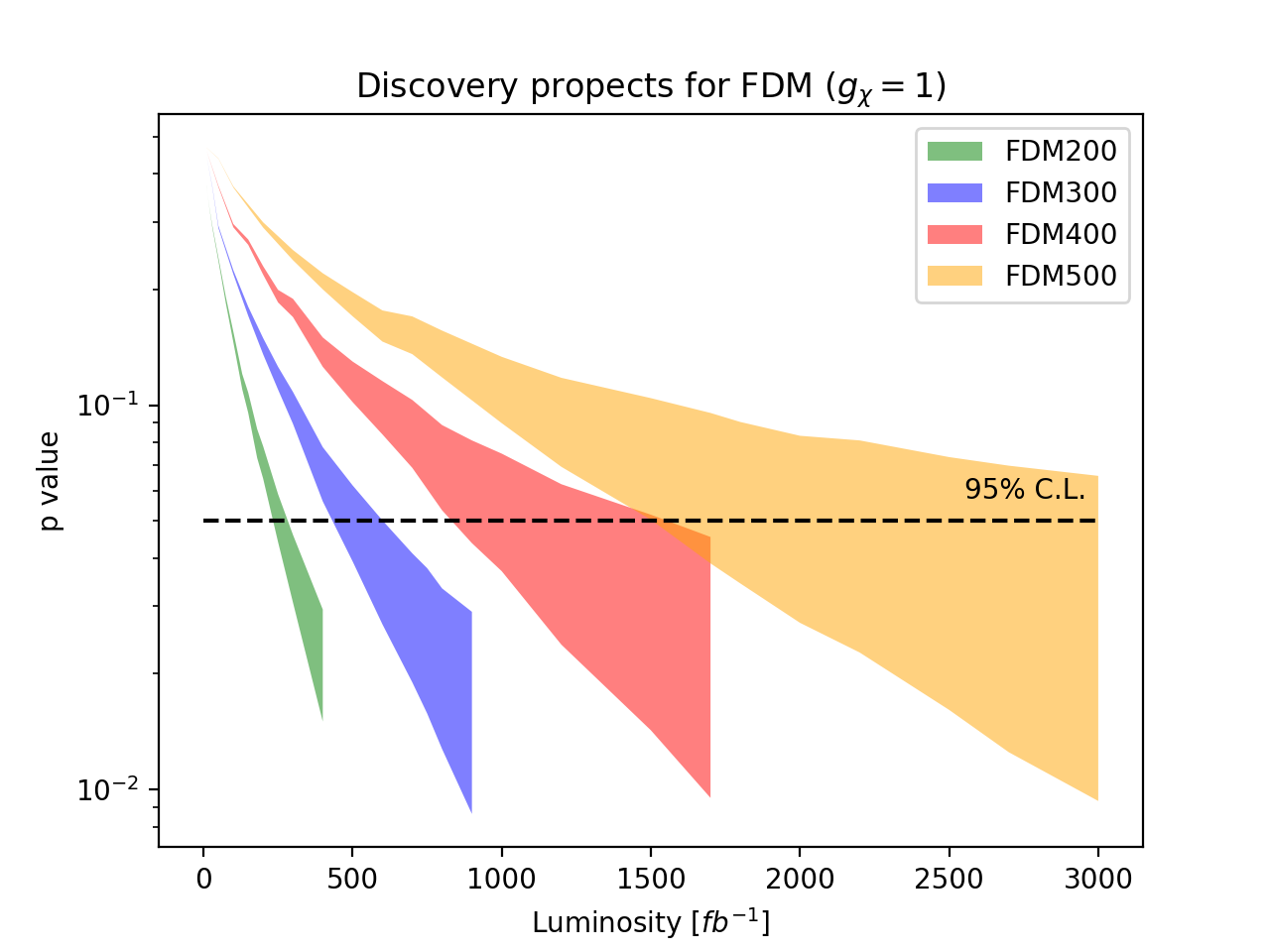}
\end{center}
\caption{\label{fig:f110} Discovery prospects of benchmark points with $g_\chi=1$ 
in FDM model at 100 TeV $pp$ collider.  The bands have the same meanings as explained in the caption of Fig.~\ref{fig:f3disc}.}
\end{figure}

Since all of our benchmark points are characterized by similar energy scale and kinematic features, 
we adopt the same analysis as has been proposed in Sec.~\ref{sec:ss} to the benchmark points with $g_\chi=1$: (1) preselection with exactly two opposite sign leptons and at least one $b$-jet; (2) $m_{\ell\ell} \notin [85,95]$ GeV; (3) $E^{\text{miss}}_T > 150$ GeV; (4) $m_{T_2}(\ell,\ell) >150$ GeV. 
The shapes of the $E^{\text{miss}}_T$ distributions after above selection requirements are used in the binned log-likelihood analysis to calculate the $p$-value for each benchmark point with respect to varying integrated luminosity. The discovery prospects are provided in Fig.~\ref{fig:f110}. The search sensitivity is improved for benchmark points with larger production cross section. Overall, all benchmark points of $g_\chi=1$ should be detectable with an integrated luminosity below $\sim 3000$ fb$^{-1}$, assuming the systematic uncertainty $\lesssim 1\%$. 
Comparing to the Fig.~\ref{fig:f3disc}, the change of prospects due to the choice of $g_\chi$ values is visible when the $H_2$ is relatively heavy ($m_{H_2}\gtrsim 400$ GeV), where the DM production through off-shell $H_2$ contribution is dominant.

\begin{figure}[htb]
\begin{center}
\includegraphics[width=0.48\textwidth]{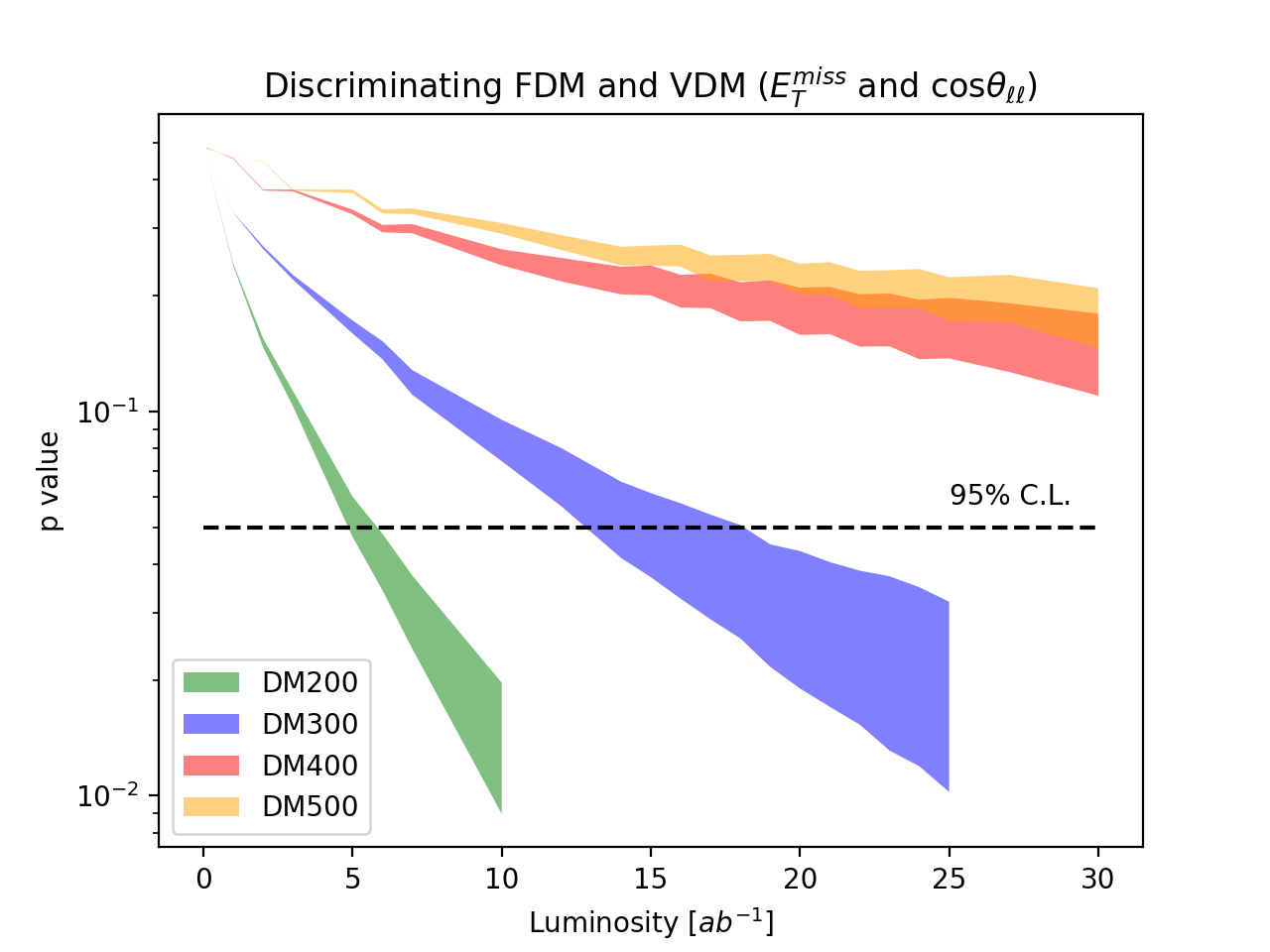}
\includegraphics[width=0.48\textwidth]{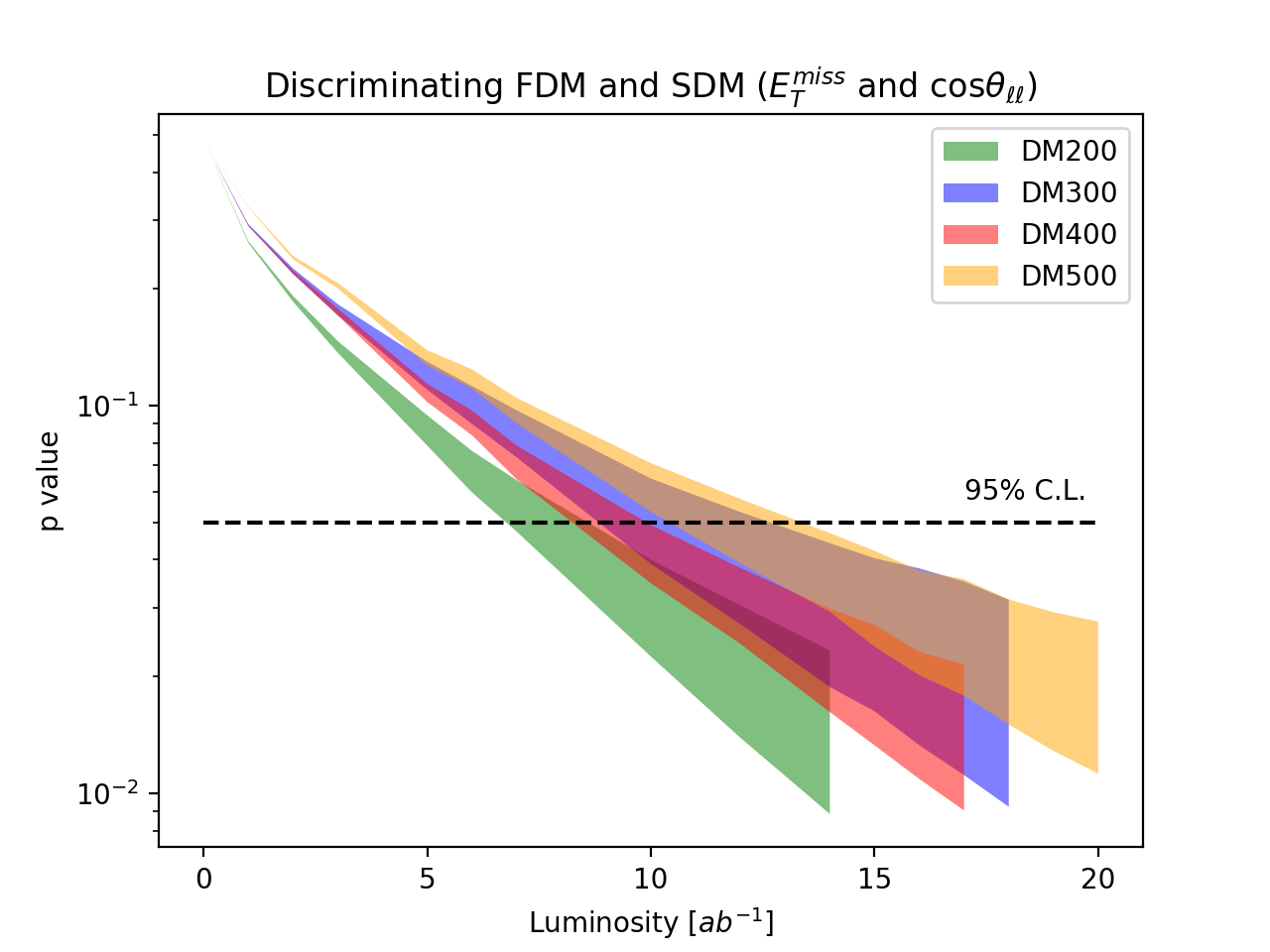}
\end{center}
\caption{\label{fig:fvs110} Spin discrimination prospects between FDM and VDM in the upper panels; between FDM and SDM in the lower panels. 
The two dimension distributions in the $\cos(\theta_{\ell \ell})$ and $E^{\text{miss}}_T$ plane are used. 
The bands have the same meanings as explained in the caption of Fig.~\ref{fig:fvs32d}.
}
\end{figure}

Finally, we consider the spin discrimination for those benchmark points with $g_\chi=1$. 
As before, the two dimensional binned log-likelihood test is performed on the distributions in the $E^{\text{miss}}_T$ and $|\cos(\theta_{\ell \ell})|$ plane. The resulting $p$-values for each case are plotted in Fig.~\ref{fig:fvs110}, where we also consider the case with systematic uncertainty of $0.5\%$. For the $g_\chi=1$ case, distinguishing between FDM and VDM will be very difficult, especially when $H_2$ is heavier.  
Because the on-shell $H_2$ production becomes dominant for a small coupling as well as 
the signal production rate gets smaller for heavier $H_2$.  We conclude that the spin discrimination is only possible when $m_{H_2} \lesssim 300$ GeV. The discrimination of FDM and SDM is relatively easier due to the intrinsic difference that FDM model has two scalar mediators while SDM model only has one. It will be possible to distinguish FDM from SDM with integrated luminosity below $\sim 15$ ab$^{-1}$ for all benchmark points. 

\section{DM phenomenology in the Higgs portal DM models and their extensions }
\label{sec:dm}

\subsection{DM phenomenology of the benchmark points}
Let us briefly discuss the DM phenomenology of our benchmark points, especially the DM relic density and direct detection constraints. We first write the complete model Lagrangains~\cite{Kamon:2017yfx} with FeynRule  and produce the CalcHEP/CompHEP~\cite{Belyaev:2012qa} model files. The model files are used by micrOMEGAs~\cite{Belanger:2008sj} to calculate the relic density and direct detection for each benchmark point. The results ($g_\chi=3$ case) are presented in Table~\ref{dmbps}. 
According to our choices of benchmark points, the DM particles dominantly annihilate into 
$WW^{(*)}$ through scalar mediator(s) in the early universe for any DM spin. 
The DM relic density for all benchmark points are below the measurement 
($\Omega h^2_0=0.1198$)~\cite{Ade:2015xua}. For FDM case, because the DM annihilation 
is $p$-wave suppressed, its relic density is larger than that of SDM and VDM. 
Comparing the rescaled DM nucleon scattering cross section (by a factor of $\frac{\Omega h^2}{0.1198}$) and the LUX constraints~\cite{Akerib:2016vxi}, we would conclude that all of our 
benchmark points should have already been excluded by the direct detection experiment. 

However, there are several ways to evade this issue. On one hand, the direct detection limits depend on assumptions about the local dark matter density and DM velocity distributions, which are 
expected to vary from the standard assumptions used in the experimental analyses~\cite{Kuhlen:2009vh,Lisanti:2010qx,Mao:2013nda,Kuhlen:2013tra}.  
Furthermore, the direct detection cross 
section depends on hadronic matrix elements which also have considerable uncertainties ~\cite{Accomando:1999eg,Ellis:2005mb}.
On the other hand, the $E^{\text{miss}}_T$ signatures at colliders could be 
generated not by the real DM candidate
that is responsible for the DM relic density of the universe, but by some heavier dark states    
that can either decay or annihilate into the proper DM candidate of the universe.  Then  the  stringent DM direct 
detection constraints would not be applicable to these heavier dark states.   In the following, 
we provide two possible scenarios which have correct relic density and evade the DM direction detection, while keeping the collider phenomenology almost the same as the benchmark points for the FDM case. The direct detection problem for VDM and SDM models can be solved in a similar way.

\begin{table}[htb]
\begin{center}
\begin{tabular}{|c|c||c|c|c|c|c|} \hline
\multicolumn{2}{|c||}{$m_{H_2}$ [GeV]}  & 200 & 300 & 400 & 500\\ \hline \hline
\multirow{ 2}{*}{FDM} & $\Omega h^2$ & $7.18 \times 10^{-3}$  & $1.18 \times 10^{-2}$ & $1.28 \times 10^{-2}$ &  $1.33 \times 10^{-2}$  \\  \cline{2-6}
& $\sigma^{\text{SI}}_p \cdot \frac{\Omega h^2}{0.1198}$ [pb] & $2.28 \times 10^{-9}$ & $1.13 \times 10^{-8}$ & $1.61 \times 10^{-8}$ & $1.87 \times 10^{-8}$   \\  \hline \hline
\multirow{ 2}{*}{VDM} & $\Omega h^2$ &  $4.78 \times 10^{-4}$ &  $1.60 \times 10^{-3}$ &  $3.05 \times 10^{-3}$ &  $4.88 \times 10^{-3}$ \\  \cline{2-6}
& $\sigma^{\text{SI}}_p \cdot \frac{\Omega h^2}{0.1198}$ [pb] & $8.44 \times 10^{-10}$ & $3.93 \times 10^{-9}$ & $5.32 \times 10^{-9}$ & $5.97 \times 10^{-9}$    \\  \hline \hline
\multirow{ 2}{*}{SDM} & $\Omega h^2$ & $2.83 \times 10^{-5}$ & $4.95 \times 10^{-5}$ & $1.04 \times 10^{-4}$ & $1.72 \times 10^{-4}$ \\  \cline{2-6}
& $\sigma^{\text{SI}}_p \cdot \frac{\Omega h^2}{0.1198}$ [pb] & $3.02 \times 10^{-9}$ & $2.94 \times 10^{-9}$ & $2.85 \times 10^{-9}$ & $2.78 \times 10^{-9}$   \\  \hline
\end{tabular}
\caption{\label{dmbps}  Relic densities and direct detection rates for benchmark points with 
$g_{\chi}=3$. }
\end{center}
\end{table}

If we choose smaller $g_\chi \sin\alpha$ and $m_\chi > m_{H_1}, m_{H_2}$, there is ample 
parameter space where FDM models provide thermal DM without violating stringent constraints from 
the direct detection experiments. However, in this case the production cross section at high energy 
collider becomes too small, and probably it is outside the reach of a future collider.

\subsection{Towards more complicated cases :  Higgs portals to excited dark states}

In a generalized case, the DM sector consists of two DM particles $\chi_1$ and $\chi_2$, where we assume $m_{\chi_2} > m_{\chi_1}$. The complete model Lagrangian is given as~\cite{ko:pre} 
\begin{align}
\mathcal{L} =&  \sum_{i=1,2} \bar{\chi}_i (i\slashed{D} - m_i -y_i S) \chi_i - [\bar{\chi}_1 (y_s + i y_p \gamma^5) S \chi_2 + h.c] \nonumber \\ 
& -\frac{1}{4} V_{\mu\nu}V^{\mu\nu}  + \frac{1}{2} D_\mu S D^\mu S - \frac{1}{2} m^2_S S^2 - V(H,S), 
\end{align}
where we have introduced an extra $U(1)_D$ dark gauge group with dark photon $V_\mu$; 
$S$ is an SM singlet complex scalar with nonzero $U(1)_D$ charge~\footnote{The $\chi_{1,2}$ in the Lagrangian are not mass eigenstates if $S$ develops VEV. We will not distinguish between mass eigenstates and gauge eigenstate in our discussion for simplicity.}.  
In the scalar potential $V(H,S)$, $S$ can develop a vacuum expectation value (VEV) providing the 
(additional) masses for the dark photon (two dark fermions). 
Also, it can mix with the SM Higgs ($H$) boson giving rise to a possible collider detection of the 
fermionic DM sector. 

In order to produce the benchmark points in the model, we require $y_2=g_\chi$, $m_{\chi_2} =80$ GeV, $m_S=\{ 200,~300,~400,~500 \}$ GeV and the scalar mixing angle $\sin \alpha =0.3$. The model with this parameter setup will generate exactly the same collider signals as discussed before. On the other hand, since $m_{\chi_2} > m_{\chi_1}$, $\chi_2$ can annihilate into $\chi_1$ while the reverse is not true at low temperature. Moreover, there could be decay channels $\chi_2 \to \chi_1 \gamma_D$, and $\chi_2 \to \chi_1 S^*(\to \chi_1 \chi_1)$. All those facts wash out the existence of the $\chi_2$ particles since the very early stage of the universe while $\chi_2$ can be copiously produced at hadron collider and leave the detector as missing transverse energy. 

$\chi_1$ particles are responsible for the relic density and astrophysics evidences of DM. In the early universe, $\chi_1$ pair can dominantly annihilate into two dark photons through t-channel process. Meanwhile, the DM direct detection constraints can be easily evaded as long as the $y_1$ coupling is sufficiently small.  More details will be presented elsewhere~\cite{ko:pre}.

\subsection{Pseudoscalar mediator  mixing with the SM Higgs}

Another simple scenario to evade the stringent DM direct detection constraints is to change the coupling form between the mediator and the DM particle, e.g., use the pseudoscalar coupling. The DM phenomenology and the collider phenomenology of the minimal FDM model with a pseudoscalar coupling have been studied for the following interaction Lagrangian~\cite{Baek:2017vzd}:
\begin{align}
\mathcal{L}^{\text{int}}_{\text{FDMSA}} &=   - i g^A_\chi (H_0 \sin \alpha + A \cos \alpha) ~ \bar{\chi} \gamma^5 \chi -(H_0 \cos \alpha - A \sin \alpha) \nonumber \\
 & \times  \left[ \sum_f \frac{m_f}{v_h} \bar{f} f - \frac{2m^2_W}{v_h} W^+_\mu W^{-\mu} - \frac{m_Z^2}{v_h} Z_\mu Z^\mu  \right],\label{langint}
\end{align}
where $H_0$ plays the role of SM Higgs and $A$ is the SM singlet scalar. 

In this model, the matrix element of the DM-nucleon scattering is proportional to the DM velocity 
\begin{align}
\mathcal{M} & \propto  \mathcal{M}_\chi  \cdot \mathcal{M}_f = -2q^i (\xi_\chi^{\dagger} \hat{S}^i \xi_\chi )\times  \nonumber \\
 & [2m_f (\xi_f^{\dagger} \xi_f) + i\frac{\mu}{m_f} \epsilon^{ijk} q^i v^{j} (\xi_f^{\dagger} \hat{S}^k \xi_f) ]~, \label{sigDD}
\end{align}
which leads to $v^2 \sim 10^{-6}$ suppression in the DM-nucleon scattering 
cross section:
\begin{align}
\sigma^{SI}_{\chi N} = \frac{2}{\pi} \frac{\mu^4}{m_\chi^2} \lambda_N^2 v^2,
\end{align}
where
\begin{align}
\lambda_N = \frac{g_\chi \sin\alpha \cos\alpha m_N}{v_h} \left(\frac{1}{m_{H_0}^2} -\frac{1}{m_A^2}\right) f_N,
\end{align}
with $N$ denoting nucleon and $f_N \approx 0.28$~\cite{Young:2009zb,Oksuzian:2012rzb,Alarcon:2011zs,Alarcon:2012nr}. 
In contrast, the $s$-wave DM annihilation is still permitted which requires the DM relic density of our benchmark points to be below the observation. This means that the DM considered in this example only constitutes a fraction of the total amount of DM sector. 

By changing the coupling from scalar to pseudoscalar, the main kinematic features of the signal at hadron collider is unaltered. But it is still possible to distinguish between those two scenarios with similar technique as adopted for spin discrimination. 
We can write the differential cross section of DM production as (Appendix~\ref{sec:app})
\begin{align}
\frac{d \sigma_{\text{FDMSA}}}{d t} & \propto  \sigma_{\text{FDMSA}}^{h^*} \times |\frac{1}{t-m^2_{H_0} + i m_{H_0} \Gamma_{H_0} } - \frac{1}{t-m^2_{A} + i m_{A} \Gamma_{A} }  |^2 \cdot 2t . \label{eq:sfdmsa}
\end{align}
Comparing to Eq.~\ref{eq:sfdm}, we can find the weight factor to be $(2t-8m^2_\chi)$
for scalar and $2t$ for pseudoscalar. 
Because $\frac{2t_2-8m^2_\chi}{2t_1-8m^2_\chi} > \frac{2t_2}{2t_1}$ for $t_2>t_1$, we expect that the $m_{\chi\chi}$ spectrum in scalar mediator model will be harder than that in pseudoscalar model. 

\begin{figure}[htb]
\begin{center}
\includegraphics[width=0.48\textwidth]{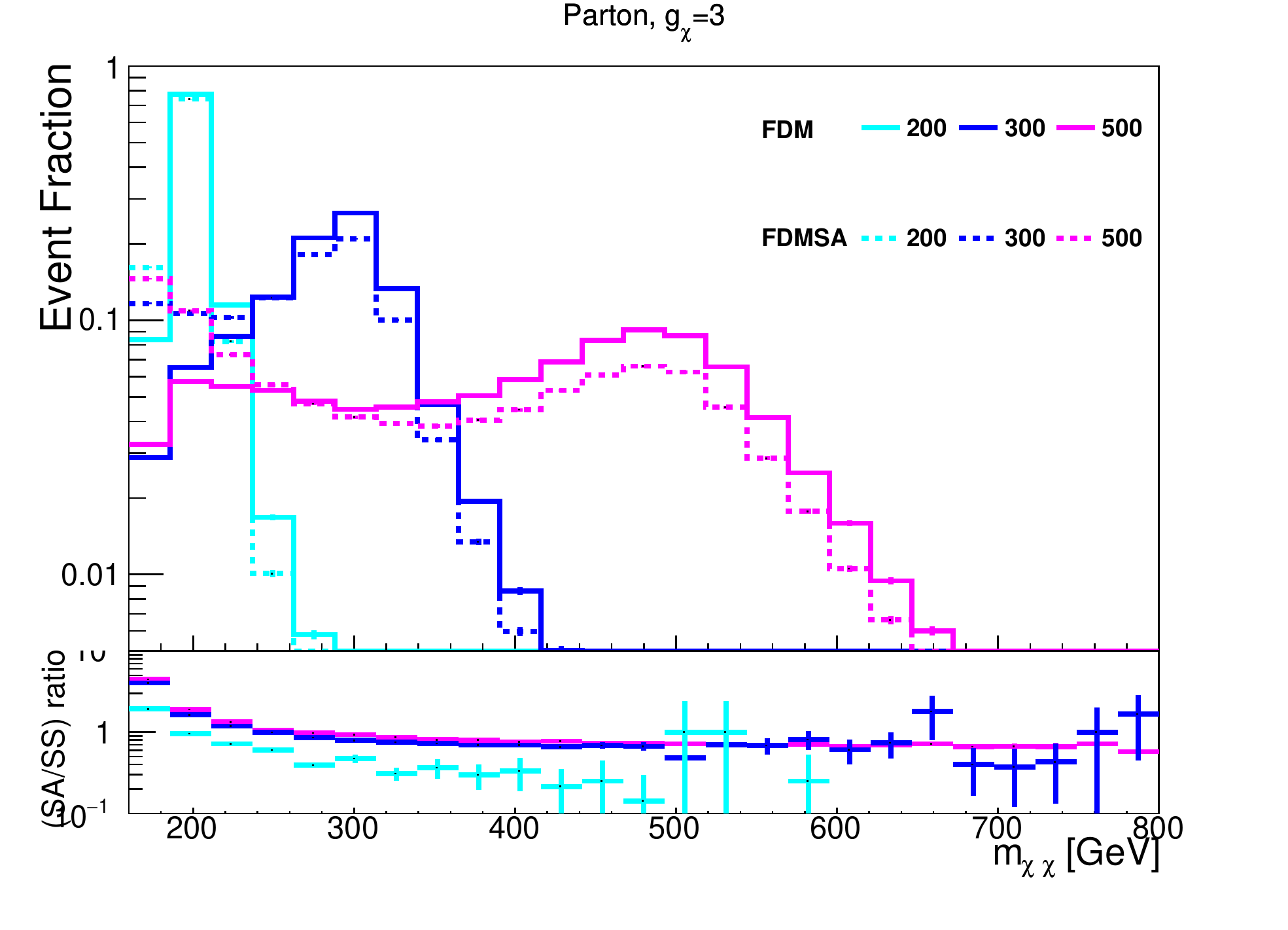}
\includegraphics[width=0.48\textwidth]{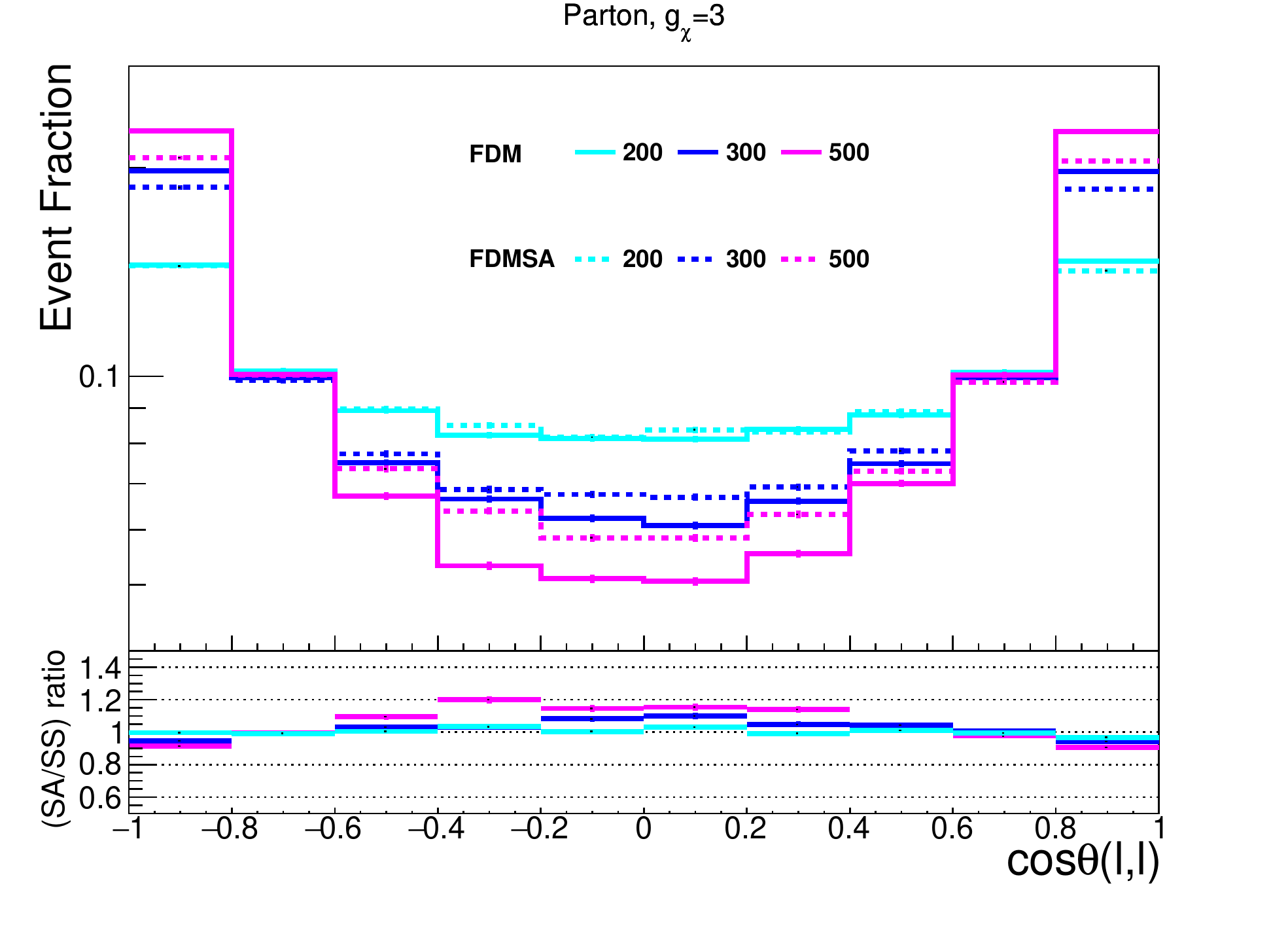}
\end{center}
\caption{\label{fig:fsa} Parton level distributions of DM pair invariant mass (left) and polar angle difference between two leptons (right) at 100 TeV $pp$ collider.  }
\end{figure}

To demonstrate the argument, we choose four benchmark points in the pseudoscalar mediator FDM model, denoted by FDMSA200, FDMSA300, FDMSA400 and FDMSA500 corresponding to those of FDM. 
Here the S/A indicates that, in this model, the mediator couples to SM fermions/DM with scalar/pseudoscalar coupling. 
The coupling $g_\chi^A$ for each benchmark point is chosen such that the decay width of A is the same as $H_2$ of the corresponding benchmark point in FDM models. The $m_{\chi\chi}$ distributions for all benchmark points are plotted in the left panel of Fig.~\ref{fig:fsa}. It can be clearly seen that the spectra of FDMSA is softer. 
This feature is inherited by the two lepton angular separation as a physical observable. The distributions of polar angle difference between the two leptons for those benchmark point are provided in the right panel of the figure.  Events with lager $m_{\chi\chi}$ will have larger angular separation between two leptons. 

\begin{figure}[htb]
\begin{center}
\includegraphics[width=0.48\textwidth]{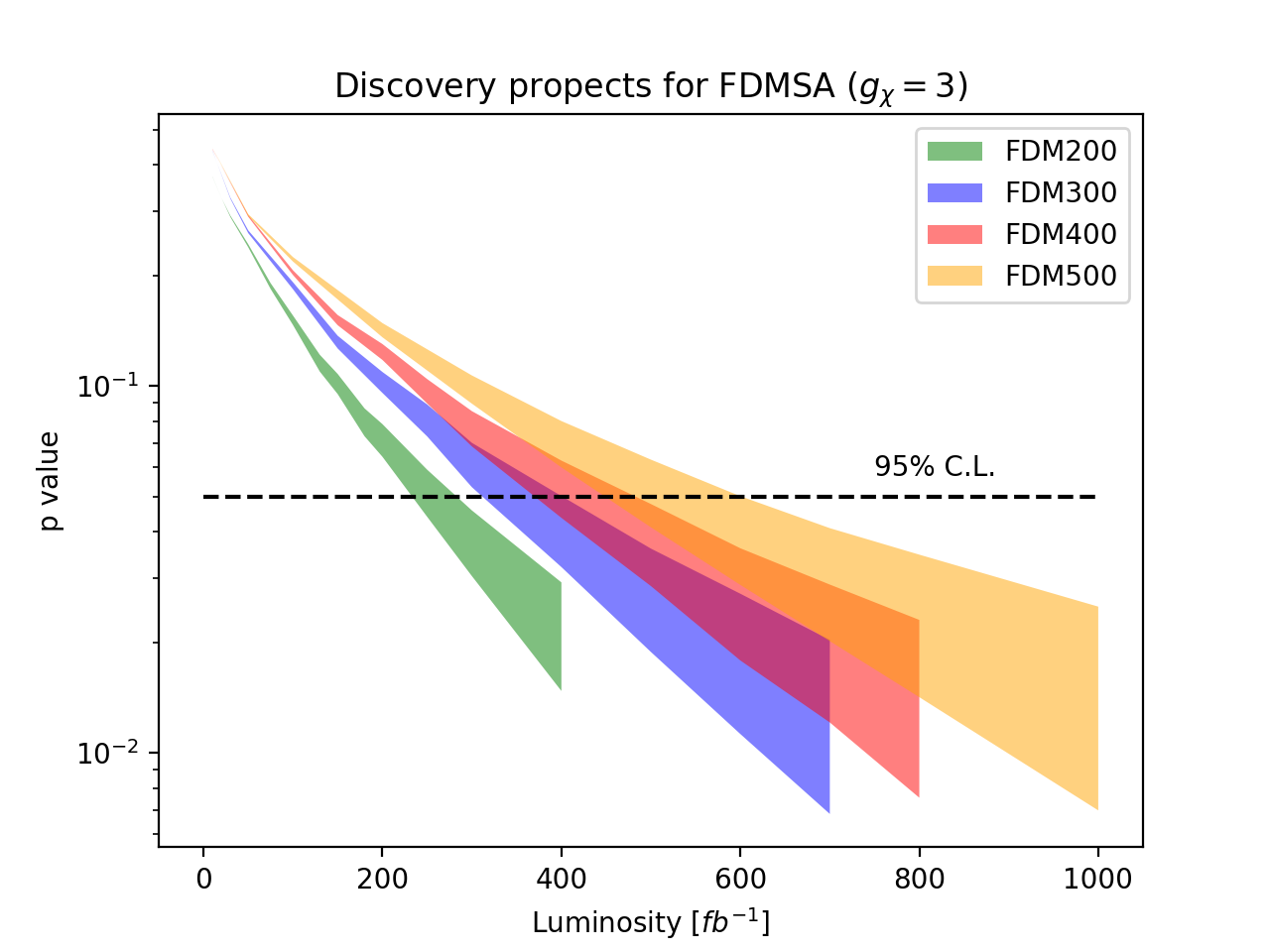}
\includegraphics[width=0.48\textwidth]{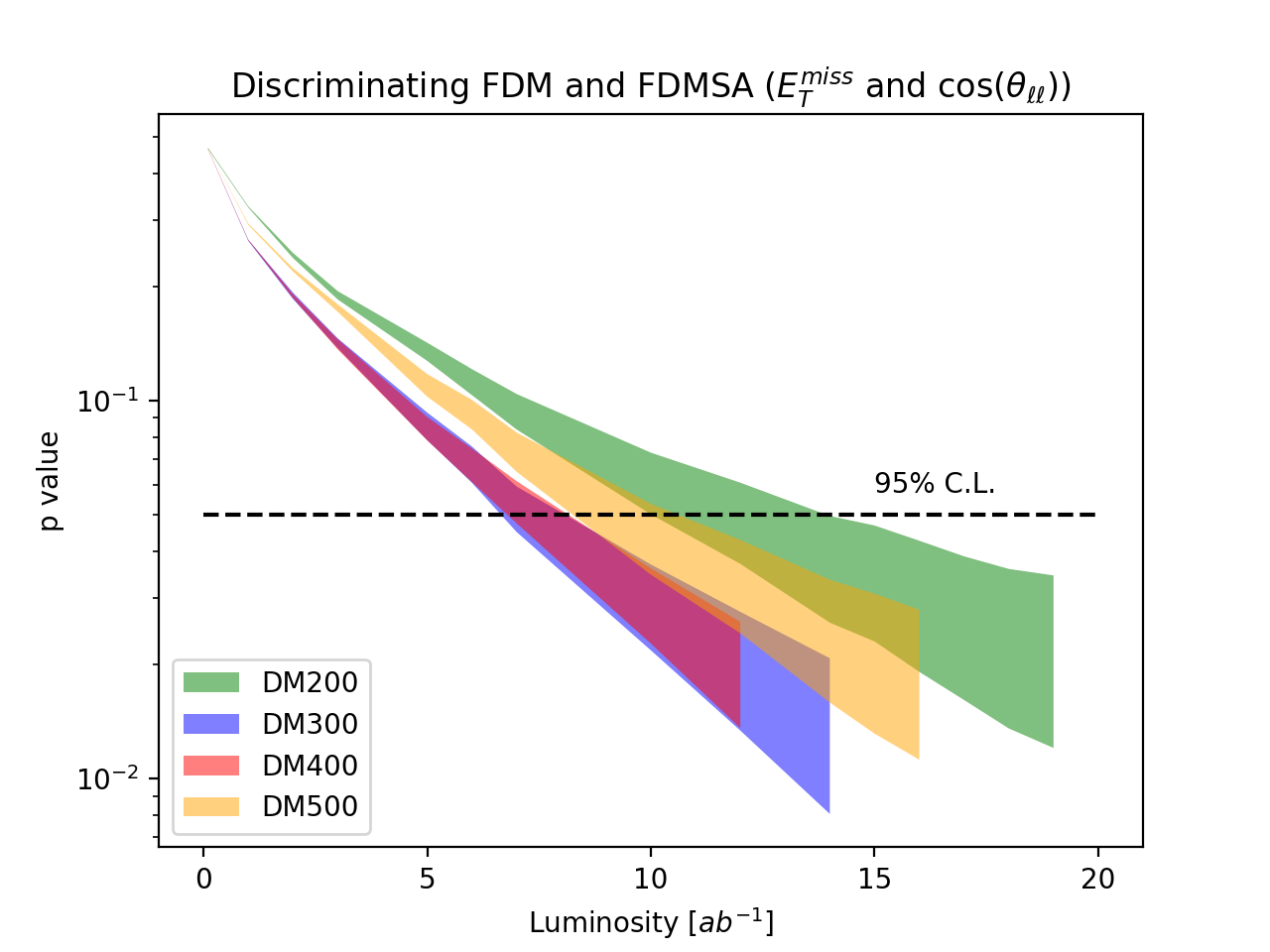}
\end{center}
\caption{\label{fig:fsadisc} Left: discovery prospects of benchmark points in FDMSA model; Right: discriminating prospects between FDM and FDMSA models. The systematic uncertainties are taken to be 1\% and 0.5\% in discovery and discrimination, respectively. }
\end{figure}

Again, we adopt the same analysis strategy as in Sec.~\ref{sec:collider} to study the discovery and discriminating (from FDM) prospects of the FDMSA model. 
The shape information of $E^{\text{miss}}_T$ distribution has been used in signal probing. The $p$-value with respect to the integrated luminosity for each FDMSA benchmark point is plotted in the left panel of Fig.~\ref{fig:fsadisc}. Similar to the FDM case, all benchmark points are probeable at 95\% C.L. for integrated luminosity below $\sim 500$ fb$^{-1}$ given a 1\% systematic uncertainty. 
In order to discriminate the FDM benchmark points from those of FDMSA, both the shapes of $E^{\text{miss}}_T$ and $\cos(\theta_{\ell \ell})$ are taking into account. The two dimensional binned log-likelihood analysis shows that the discrimination can be made with an integrated luminosity of around 15 ab$^{-1}$ for all benchmark points if the systematic uncertainty can be controlled at 0.5\% level. 

\section{Summary}
\label{sec:concl}

In this paper, we have investigated prospects of the DM discovery and its spin discrimination at a 100 TeV $pp$ collider for the Higgs portal DM models and their extensions 
with the $t\bar{t}+$DM associated production in the dileptonic channel. Kinematic variable of dilepton invariant mass $m_{\ell\ell}$, missing transverse energy  $E^{\text{miss}}_T$ and stransverse mass of the two leptons $m_{T_2} (\ell, \ell)$ are used in our cut-and-count analysis for the first stage of signal and background discrimination. Especially, the $m_{T_2} (\ell, \ell)$ is found to be useful in suppressing the SM $t\bar{t}$ background. The shape information of the $E^{\text{miss}}_T$ is used further by one dimensional binned log-likelihood test to estimate the signal discovery prospects. We find that our benchmark points can be probed at a future $pp$ collider with an integrated luminosity 
below $\sim$1 ab$^{-1}$, assuming the systematic uncertainty can be controlled at 1\% level. 

The models with different DM spins are predicting different distributions in the variable 
$t\equiv m_{DD}^2$. Even though the $t$ variable itself is not an observable at hadron collider, 
its feature can be reflected in the angular separation between recoiling two top quarks. 
We adopt a two dimensional binned log-likelihood analysis on the distributions of missing transverse 
energy and two leptons (from top quark decay) polar angle difference for different signals plus backgrounds for the DM spin discrimination. Our study shows that the DM spin discrimination is possible at a future 100 TeV collider with an integrated luminosity below a few $\mathcal{O}(10)$ ab$^{-1}$ for most cases if the systematic uncertainty can be controlled at $\sim 0.5\%$ level. 
By applying the same analysis to more general cases with smaller couplings ($g_\chi=1$), our findings do not change much except that the DM spin discrimination become very difficult when the coupling is small and the mediator ($H_2$) is heavy.  

Finally, we discuss the DM phenomenology of our benchmark points, where relic densities are well below the measurement and which are challenged by DM direct detection experiments. Two possible solutions are proposed to avoid these issues: (1) extending the DM sector where the DM particle of interest at collider is not the same as the DM particle from astrophysics observation; (2) modifying the DM coupling such that the non-relativistic DM-nucleon scattering is suppressed, i.e. using pseudoscalar coupling between the DM and the mediator. Discrimination between the scalar and the pseudoscalar couplings is shown to be quite promising 
at a future 100 TeV $pp$ collider.

\section*{Acknowledgement}

We are grateful to Tathagata Ghosh for discussions on the related issues.
This work is supported in part by National Research Foundation of Korea (NRF) Research Grant NRF-2015R1A2A1A05001869 (PK, JL), by the NRF grant funded by the Korea government 
(MSIP) (No. 2009-0083526) through Korea Neutrino Research Center at Seoul National University (PK), 
by DOE Grant DE-SC0010813 (BD,TK) and by Qatar National Research Fund under project NPRP 9-328-1-066 (TK).

\phantomsection
\addcontentsline{toc}{section}{References}
\bibliographystyle{jhep}
\bibliography{DMFCC}
\appendix
\section{Matrix element calculation}
\label{sec:app}
In this appendix, we calculate the spin summed matrix element square for each case that is discussed in current work. 
\subsection{Fermion DM with scalar interaction}
Assuming the mediator is propagating along the $z-$axis without loss of generality, 
the wave function for the outgoing DM and anti-DM are given by:
\begin{align}
u^p_+ &= \frac{1}{\sqrt{2 |\vec{p}| (|\vec{p}| + p_z)}} \{ \omega^p_{-} (|\vec{p}| + p_z ),~ \omega^p_- p_x, ~ \omega^p_+ (|\vec{p}| + p_z), ~\omega^p_+ p_x \} ^T , \label{up} \\ 
u^p_- &=  \frac{1}{\sqrt{2 |\vec{p}| (|\vec{p}| + p_z)}} \{ \omega^p_+ p_x, ~\omega^p_+ (|\vec{p}| + p_z), ~- \omega^p_- p_x, ~\omega^p_- (|\vec{p}| + p_z) \}^T , \label{um} \\
v^q_+ &=  \frac{1}{\sqrt{2 |\vec{q}| (|\vec{q}| + q_z)}} \{ \omega^q_+ q_x, ~ -\omega^q_+ (|\vec{q}| + q_z), ~ -\omega^q_- q_x, ~ \omega^q_- (|\vec{q}| + q_z)  \}^T, \label{vp} \\
v^q_- &=  \frac{1}{\sqrt{2 |\vec{q}| (|\vec{q}| + q_z)}}  \{ \omega^q_- (|\vec{q}| + q_z), ~ \omega^q_- q_x ,~ -\omega^q_+ (|\vec{q}| + q_z), ~-\omega^q_+ q_x \} ^T, \label{vm}
\end{align}
where $\omega_{\pm} = \sqrt{E \pm |\vec{p}|}$; $p$ and $q$ are the four momentum of DM and anti-DM respectively. 
The spin summed matrix element square is
\begin{align}
\sum |\bar{u}(p) v(q)|^2  & =  \frac{(p_x q_x + |\vec{p}| (|\vec{q}| + q_z ) + p_z (|\vec{q}| +q_z) )^2 }{2 |\vec{p}|  |\vec{q}|  (|\vec{p}| + p_z) (|\vec{q}| + q_z) } (\omega^p_+ \omega^q_-  - \omega^p_- \omega^q_+ )^2  \nonumber \\ 
& +   \frac{( q_x (|\vec{p}| + p_z) - p_x (|\vec{q}| +q_z) )^2}{2 |\vec{p}|  |\vec{q}|  (|\vec{p}| + p_z) (|\vec{q}| + q_z) } (\omega^p_- \omega^q_- - \omega^p_+ \omega^q_+). 
\end{align}
If one take the DM momenta  in the rest frame of the mediator with mass given by $\sqrt{t}=m_{DD}$:
\begin{align}
p &= \{ \frac{\sqrt{t}}{2} , \sqrt{\frac{t}{4} - m^2_D } \sin \theta, 0 ,   \sqrt{\frac{t}{4} - m^2_D }  \cos \theta \}, \\
q &= \{ \frac{\sqrt{t}}{2} , - \sqrt{\frac{t}{4} - m^2_D } \sin \theta, 0 ,   - \sqrt{\frac{t}{4} - m^2_D }  \cos \theta \},
\end{align}
we can obtain 
\begin{align}
\sum |\bar{u}(p) v(q)|^2 = 2t - 8 m^2_D. 
\end{align}

\subsection{Fermion DM with pseudoscalar interaction}
The wave functions for DM and anti-DM are the same as in Eqs.~\ref{up}-\ref{vm}. 
The matrix element square is 
\begin{align}
\sum |\bar{u}(p) \gamma^5 v(q)|^2  & =  \frac{(p_x q_x + |\vec{p}| (|\vec{q}| + q_z ) + p_z (|\vec{q}| +q_z) )^2 }{2 |\vec{p}|  |\vec{q}|  (|\vec{p}| + p_z) (|\vec{q}| + q_z) } (\omega^p_+ \omega^q_-  + \omega^p_- \omega^q_+ )^2  \nonumber \\ 
& +   \frac{( q_x (|\vec{p}| + p_z) - p_x (|\vec{q}| +q_z) )^2}{2 |\vec{p}|  |\vec{q}|  (|\vec{p}| + p_z) (|\vec{q}| + q_z) } (\omega^p_- \omega^q_- + \omega^p_+ \omega^q_+) \nonumber \\
 &\myeq ~~~~ 2 t. 
\end{align}

\subsection{Vector DM with scalar interaction}

Assuming the mediator is propagating along $z-$axis, and two VDM momenta are
\begin{align}
k^\mu_1 &= \{ E_1, |k_1| \sin \theta_1, 0, |k_1| \cos \theta_1 \} ,\\
k^\mu_2 &= \{ E_2, |k_2| \sin \theta_2, 0, |k_2| \cos \theta_2 \} ,
\end{align}
the three independent polarization vectors can be written as 
\begin{align}
\epsilon(k_i, 1) & = \{ 0, \cos \theta_i, 0, -\sin \theta_i \} , \\
\epsilon(k_i, 2) & = \{ 0,0,1,0 \} , \\
\epsilon(k_i, 3) & = \{ \frac{k_i}{m_D}, \frac{e_i}{m_D} \sin \theta_i, 0, \frac{e_i}{m_D} \cos \theta_i \} . 
\end{align}

As a result, the matrix element square with all DM polarization summed over is given by
\begin{align}
\sum_{i,j} |g_{\mu\nu} \epsilon^{\mu *}(k_1,i) \epsilon^\nu (k_2,j) |^2 &= 1 + \cos^2(\theta_1 - \theta_2) + \frac{\sin^2(\theta_1 - \theta_2)}{m^2_D} (E_1^2 - E^2_2) \nonumber \\
& + \frac{(k_1 k_2 - E_1 E_2 \cos(\theta_1 - \theta_2))^2}{m^4_D}  \\
& \myeq ~~~~ 2+ \frac{ (t- 2 m^2_D)^2 }{4m_D^4}
\end{align}

\end{document}